\documentclass[english,prb,reprint,superscriptaddress]{revtex4-1}
\usepackage[latin9]{inputenc}
\usepackage{color}
\usepackage{textcomp}
\usepackage{amsmath}
\usepackage{amssymb}
\usepackage{graphicx}
\usepackage{subscript}

\makeatletter

\DeclareFontEncoding{LGR}{}{}
\DeclareRobustCommand{\greektext}{%
  \fontencoding{LGR}\selectfont\def\encodingdefault{LGR}}
\DeclareRobustCommand{\textgreek}[1]{\leavevmode{\greektext #1}}
\ProvideTextCommand{\~}{LGR}[1]{\char126#1}

\providecommand{\tabularnewline}{\\}

\usepackage{textcomp}
\providecommand{\tabularnewline}{\\}

\usepackage{array}

\usepackage{babel}

\makeatother

\usepackage{babel}
\begin{document}
\title{The two-dimensional electron gas of the In\textsubscript{2}O\textsubscript{3}
surface: Enhanced thermopower, electrical transport properties, and
its reduction by adsorbates or compensating acceptor doping}
\author{Alexandra Papadogianni}
\affiliation{Paul-Drude-Institut für Festkörperelektronik, Leibniz-Institut im
Forschungsverbund Berlin e.V., Hausvogteiplatz 5--7, D--10117 Berlin,
Germany}
\author{Julius Rombach}
\affiliation{Paul-Drude-Institut für Festkörperelektronik, Leibniz-Institut im
Forschungsverbund Berlin e.V., Hausvogteiplatz 5--7, D--10117 Berlin,
Germany}
\author{Theresa Berthold}
\affiliation{Institut für Physik und Institut für Mikro- und Nanotechnologien,
Technische Universität Ilmenau, PF 100565, D--98684 Ilmenau, Germany}
\author{Vladimir Polyakov}
\affiliation{Fraunhofer-Institut für Angewandte Festkörperphysik, Tullastraße 72,
D-79108 Freiburg, Germany}
\author{Stefan Krischok}
\affiliation{Institut für Physik und Institut für Mikro- und Nanotechnologien,
Technische Universität Ilmenau, PF 100565, D--98684 Ilmenau, Germany}
\author{Marcel Himmerlich}
\affiliation{Institut für Physik und Institut für Mikro- und Nanotechnologien,
Technische Universität Ilmenau, PF 100565, D--98684 Ilmenau, Germany}
\affiliation{CERN, European Organization for Nuclear Research, 1211 Meyrin, Switzerland}
\author{Oliver Bierwagen}
\affiliation{Paul-Drude-Institut für Festkörperelektronik, Leibniz-Institut im
Forschungsverbund Berlin e.V., Hausvogteiplatz 5--7, D--10117 Berlin,
Germany}
\begin{abstract}
In\textsubscript{2}O\textsubscript{3} is an \textit{n}-type transparent
semiconducting oxide possessing a surface electron accumulation layer
(SEAL) like several other relevant semiconductors, such as InAs, InN,
SnO\textsubscript{2}, and ZnO. Even though the SEAL is within the
core of the application of In\textsubscript{2}O\textsubscript{3}
in conductometric gas sensors, a consistent set of transport properties
of this two-dimensional electron gas (2DEG) is missing in the present
literature. To this end, we investigate high quality single-crystalline
as well as textured doped and undoped In\textsubscript{2}O\textsubscript{3}(111)
films grown by plasma-assisted molecular beam epitaxy (PA-MBE) to
extract transport properties of the SEAL by means of Hall effect measurements
at room temperature while controlling the oxygen adsorbate coverage
via illumination. The resulting sheet electron concentration and mobility
of the SEAL are $\approx1.5\times10^{13}$~cm$^{-2}$ and $\approx150$~cm$^{2}$/Vs,
respectively, both of which get strongly reduced by oxygen-related
surface adsorbates from the ambient air. Our transport measurements
further demonstrate a systematic reduction of the SEAL by doping In$_{2}$O$_{3}$
with the deep compensating bulk acceptors Ni or Mg. This finding is
supported by X-ray photoelectron spectroscopy (XPS) measurements of
the surface band bending and SEAL electron emission. Quantitative
analyses of these XPS results using self-consistent, coupled Schrödinger--Poisson
calculations indicate the simultaneous formation of compensating bulk
donor defects (likely oxygen vacancies) which almost completely compensate
the bulk acceptors. Finally, an enhancement of the thermopower by
reduced dimensionality is demonstrated in In$_{2}$O$_{3}$: Seebeck
coefficient measurements of the surface 2DEG with partially reduced
sheet electron concentrations between $3\times10^{12}$ and $7\times10^{12}$~cm$^{-2}$
(corresponding average volume electron concentration between $1\times10^{19}$
and $2.3\times10^{19}$~cm$^{-3}$) indicate a value enhanced by
$\approx80$\,\% compared to that of bulk Sn-doped In\textsubscript{2}O\textsubscript{3}
with comparable volume electron concentration. 
\end{abstract}
\date{\today}
\maketitle

\section{Introduction}

Indium oxide (In$_{2}$O$_{3}$) is a transparent semiconducting oxide,
which exhibits inherent \textit{n}-type conductivity, commonly referred
to as unintentional doping (UID). Like the related or SnO$_{2}$\citep{Nagata2011},
In$_{2}$O$_{3}$ possesses a surface electron accumulation layer
(SEAL)\citep{King_2009_In2O3}, that lies within the core of In$_{2}$O$_{3}$-based
conductometric gas sensors for oxygen species.\citep{rombach_SAB-2016}
Along with this, In$_{2}$O$_{3}$ typically finds applications as
a transparent contact in optoelectronic devices, mostly in its highly
Sn-doped form, known as ITO\citep{chae2001,tiwari2004,tsai_2016},
which can reach electron concentrations as high as $10^{21}\thinspace\mathrm{cm^{-3}}$.
This particular application of In$_{2}$O$_{3}$ further benefits
from the existence of the SEAL, which favors the formation of Ohmic
contacts. This property, nevertheless, indicates that the formation
of Schottky contacts---required for several other applications---is
hindered by the existence of the SEAL, even for high work function
metals like Pt\citep{michel_acsami.9b06455,vonwenckstern_schottky2014}.
Tunability of the SEAL is, hence, necessary to both unlock the entire
spectrum of potential device applications of In$_{2}$O$_{3}$ and
tune its gas sensitivity.

Using X-ray photoelectron spectroscopy (XPS) measurements \citet{king_SEAL-CNL-In2O3_2008}
demonstrated the existence of a few-nm thick electron accumulation
layer at the surface of In$_{2}$O$_{3}$ by a downward band bending
at the surface of undoped single-crystalline films, in contradiction
to previous investigations reporting a surface depletion.\citep{klein_2000_surface_depletion,Gassenbauer_2006}
This discrepancy mainly arose due the difference in the assumed fundamental
band gap of In$_{2}$O$_{3}$ required for the interpretation of the
XPS results: While the optical bandgap of $\approx$3.7~eV has been
assumed to equal the fundamental one by the authors of Ref.\,\onlinecite{klein_2000_surface_depletion,Gassenbauer_2006},
the authors of Ref.~\onlinecite{king_SEAL-CNL-In2O3_2008} have assumed
a dipole-forbidden, fundamental band gap of $\approx2.6$~eV ---
in agreement with state-of-the art \textit{ab-initio} theory combined
with bulk and surface sensitive XPS measurements.\citep{Walsh2008}
A general explanation for the existence of the In$_{2}$O$_{3}$ SEAL
has been given within the context of the charge neutrality level (CNL),
also known as branch point energy. Defect states at the CNL acquire
their weight equally from the valence and conduction bands,\citep{tersoff_schottky-barriers_1984}
essentially rendering the CNL a demarcation between donor- and acceptorlike
defect states. In contrast to most other semiconductors, in the case
of In$_{2}$O$_{3}$ the CNL lies within the conduction band,\citep{king_SEAL-CNL-In2O3_2008,schleife_2019}
due to its particular bulk band structure, with a very prominent,
low lying conduction band minimum (CBM) at the \textit{$\Gamma$}-point
and an almost flat valence band. Donorlike states at the surface of
In\textsubscript{2}O\textsubscript{3} pin the surface Fermi level,
$E_{\mathrm{F}}$, slightly below the CNL, causing a downward bending
of the conduction and valence bands. Breaking of the translational
symmetry of the bulk can give rise to such donorlike surface states.\citep{moenchw_2001_surfaces-interfaces}
Besides that, for films exposed to the ambient, the enhanced conductivity
of the surface has also been attributed to adsorbates attaching to
it; an effect not observed for films that have undergone \textit{in
situ} cleavage of the surface. \citep{Nazarzahdemoafi-2016} The microscopic
origin of the SEAL, has been further associated with surface oxygen
vacancies acting as doubly ionized shallow donors $V_{\mathrm{O}}^{2+}$\citep{zhang_in2o3_Vo-2013}
and their strongly reduced defect formation energy.\citep{Walsh2011b}
Finally, surface In adatoms, which are energetically favored over
$V_{\mathrm{O}}$,\citep{Wagner_10.1002/admi.201400289} can also
act as shallow donors\citep{Davies_10.1021/acs.jpcc.8b08623} and
have been experimentally demonstrated on the In\textsubscript{2}O\textsubscript{3}(111)
surface after a reducing surface preparation {[}annealing at 300--500\,°C
in ultra-high vacuum (UHV){]}.\citep{Wagner_10.1002/admi.201400289}

\begin{figure*}
\centering\includegraphics[width=14cm]{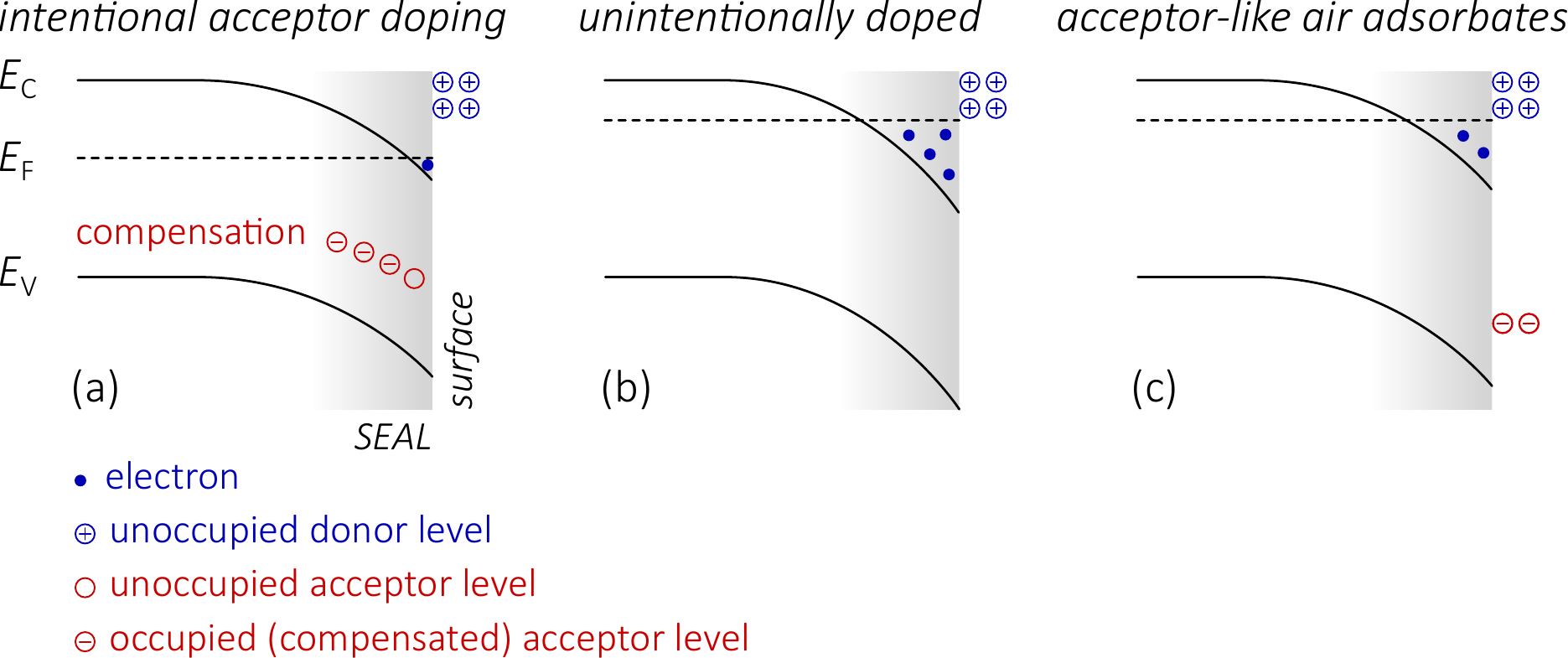}

\caption{Schematic representation of the band alignment for (a) intentional
doping with bulk acceptors and the corresponding compensation of the
SEAL, (b) unintentionally doped In\protect\textsubscript{2}O\protect\textsubscript{3},
and (c) unintentional compensation of the SEAL due to adsorption of
acceptorlike oxidizing species. The difference $(N_{D,S}^{+}-N_{A,S}^{-})$
of the 2D concentration of charged surface donors and acceptors provides
a net surface charge $N_{SS}$.\label{fig:doping-levels}}
\end{figure*}

For 111-oriented films grown by plasma-assisted molecular beam epitaxy
(PA-MBE), like the ones studied in the current work, the surface electron
concentration has been shown to have a peak value at $8\times10^{19}\,\mathrm{cm^{-3}}$\citep{King_2009_In2O3},
similar to the results from Schrödinger--Poisson modeling on the
SEAL of melt-grown bulk In\textsubscript{2}O\textsubscript{3} single
crystal studied in Ref.~\onlinecite{Nagata_2019}. Moreover, angle-resolved
photoelectron spectroscopy (ARPES) measurements have confirmed the
two-dimensional electron gas (2DEG) nature of the SEAL with sheet
electron concentration of $4\times10^{13}\,\mathrm{cm^{-2}}$ after
surface preparation at high temperature (by repeated cycles of Ar$^{+}$
sputtering (500\,eV) and annealing at 600\,°C in UHV for 1~h)\citep{zhang_in2o3_Vo-2013}
and $2\times10^{13}\,\mathrm{cm^{-2}}$ after surface preparation
at intermediate temperature (annealing at 300\,°C in UHV for $\approx15$~min).\citep{jovic_10.1002/smll.201903321}
Both of the surface preparations employed within those studies are
prone to reduce the surface---that is increase the concentration
of surface $V_{\mathrm{O}}$ or In adatoms acting as surface donors
and, thus, result in a stronger SEAL (i.e. with a higher electron
concentration) compared to that of an unprepared sample.

Significant reduction of a SEAL by compensating bulk acceptors has
been previously demonstrated in InN by Mg acceptor doping.\citep{linhart_Mg-dop_InN_SEAL}
Previous studies\citep{papadogianni_In2O3:Ni_YSZ,bierwagen2012_Mg-dop}
have shown that acceptors like Ni and Mg have a compensating effect
on the bulk electron transport of In\textsubscript{2}O\textsubscript{3}.
This effect, however, is revealed after an additional annealing of
the material in oxygen, which has been explained by overcompensation
of the added acceptors due to the simultaneous formation of donorlike
point defects---most likely $V_{\mathrm{O}}$---during growth:\citep{papadogianni_In2O3:Ni_YSZ,bierwagen2012_Mg-dop}
The addition of acceptor elements lowers the Fermi energy and, according
to Refs.~\onlinecite{lany2007,Limpijumnong_PRB_2009}, this reduces
the formation energy of $V_{\mathrm{O}}$---thus promoting their
incorporation into the crystal lattice. Studies regarding the position
of the bulk donor levels in the band gap associated with oxygen vacancies
have been rather inconclusive, with some works indicating $V_{\mathrm{O}}$
to have deep donor levels\citep{lany2007,Limpijumnong_PRB_2009} and
others to potentially be shallow donors\citep{Agoston2009,Buckeridge_PhysRevMaterials.2.054604,chatratin_PhysRevMaterials.3.074604}.\textcolor{black}{{}
Whether the annealing completely removed the doping-induced }$V_{\mathrm{O}}$
could not be clarified in Refs.~\onlinecite{bierwagen2012_Mg-dop,papadogianni_In2O3:Ni_YSZ}.
\textcolor{black}{Fig.~\ref{fig:doping-levels}~(a) shows schematically
the effect of bulk acceptor doping on the position of the $E_{\mathrm{F}}$
and the band alignment, with emphasis on its impact on the SEAL, assuming
neither spontaneous formation of compensating donors nor their removal
with a treatment such as oxygen annealing. For comparison, Fig.~\ref{fig:doping-levels}~(b)
shows the position of the $E_{\mathrm{F}}$ and band alignment in
a UID }In\textsubscript{2}O\textsubscript{3}.

Early studies on the conductivity of In\textsubscript{2}O\textsubscript{3}
at elevated temperatures have already documented its dependence on
the oxygen content of the sample environment\citep{Rupprecht1954},
which is the basis of its application as the active material in conductometric
gas sensors. At sufficiently low temperatures, that preclude oxygen
diffusion in the lattice, this sensing behavior is related to the
surface-acceptor role of adsorbed oxygen species that can reduce the
SEAL by electron transfer.\citep{Kim2013,Berthold_PSSB_2018}\textcolor{black}{{}
The effect of acceptorlike air adsorbates on the band banding and,
hence, occupation of the SEAL is schematically shown in Fig.~\ref{fig:doping-levels}~(c).}
Such gas sensors are typically (re)activated by heating the sensing
material at elevated temperatures (typically a few hundred °\,C).
Efforts towards a more energy-efficient solution have demonstrated
In\textsubscript{2}O\textsubscript{3} gas-sensors operating at room
temperature reactivated by ultraviolet (UV) light induced photoreduction
\citep{Xirouchaki1996,Imai1999,wang2007_O3sensor}. During photoreduction
the illumination forces the desorption of the negatively charged oxygen
adsorbates\citep{Wang2011} through recombination with the photogenerated
holes while the photogenerated electrons remain in the In$_{2}$O$_{3}$.\citep{Kind2002}
The SEAL sheet conductivity of PA-MBE grown In$_{2}$O$_{3}$ films
in air has been reported to be $3\times10^{-4}\,\mathrm{S}$ in the
photoreduced stationary state (under UV illumination)\citep{rombach_SAB-2016}
and below $2.2\times10^{-5}\,\mathrm{S}$ with oxygen adsorbates (i.e.,
without illumination).\citep{bierwagen2011PLOX} A reduction of the
SEAL in those films by oxygen adsorbates has been independently demonstrated
by conductance and XPS measurements.\citep{rombach_SAB-2016,Berthold_PSSB_2018}
However, these earlier works do not provide any information concerning
the actual electron concentration at the In\textsubscript{2}O\textsubscript{3}
surface.

There have, thus, been no reports regarding the full set of the SEAL
transport properties (sheet conductivity and sheet electron concentration)
measured with a single technique, after a defined surface treatment,
and in a defined environment---as has long been accomplished for
ZnO\citep{Grinshpan_PhysRevB.19.1098}, for instance. Furthermore,
there is no information in the literature concerning the thermoelectric
properties of the SEAL. In addition to the strong interest in discovering
and understanding the thermoelectric transport properties and mechanisms
of such 2DEGs, knowledge of the SEAL properties is necessary for the
application aspect of the material, as it enables controllable fine-tuning
of the (thermo)electrical behavior of the In\textsubscript{2}O\textsubscript{3}
surface.

The current work consistently determines the surface transport properties
of In\textsubscript{2}O\textsubscript{3} and demonstrates the intentional
and controllable reduction of the sheet electron concentration at
the surface of In\textsubscript{2}O\textsubscript{3} by incorporation
of the compensating bulk acceptors Ni and Mg or by oxygen surface
adsorbates. This is accomplished with the combination of Hall effect
transport measurements (with and without UV illumination) and X-ray
photoelectron spectroscopy. Supporting self-consistent Schrödinger--Poisson
calculations reveal a close compensation of the bulk acceptors by
oxygen vacancies even after annealing the samples in oxygen. Finally,
the thermoelectric properties of the surface electron accumulation
layer are investigated by Seebeck coefficient measurements. As previously
demonstrated for ZnO\citep{Shimizu6438}, the 2DEG at the In\textsubscript{2}O\textsubscript{3}
surface is shown to also exhibit an increased thermopower in comparison
bulk Sn-doped films with comparable volume electron concentration.

\section{Experiment}

For the purposes of this study, high quality (111)-oriented In\textsubscript{2}O\textsubscript{3}
has been synthesized by PA-MBE. Single-crystalline UID and Ni-doped
films have been grown on quarters of 2'' insulating ZrO\textsubscript{2}:Y
(YSZ) (111) substrates, whereas full 2'' Al\textsubscript{2}O\textsubscript{3}
(0001) (\textit{c}-plane Al\textsubscript{2}O\textsubscript{3})
substrates have been employed for the growth of UID and Mg-doped textured
films. After growth, all samples have been further cleaved into smaller
pieces with a size of approximately $5\times5\,\mathrm{mm^{2}}$.
The total thickness of the films ranges between 350--500~nm. Further
details on the growth of the studied samples are reported in Ref.~\onlinecite{papadogianni_In2O3:Ni_YSZ}
(single-crystalline) and Ref.~\onlinecite{rombach_SAB-2016} (textured).

In order to largely remove compensating donors, all samples under
study have been annealed in oxygen within a rapid thermal annealing
(RTA) system at 800\,°C at atmospheric pressure for 60~s. The undoped
samples have also been annealed in oxygen to serve as references with
comparable characteristics.

For reference measurements, an oxygen plasma treatment of the surface
at room temperature was performed in a 13.56~MHz inductively coupled
plasma (ICP) reactive ion-etching (RIE) system (Samco Inc., RIE-400iP;
process pressure, 0.025~mbar; oxygen flow, 10 standard cubic centimeters
per minute; ICP power, 100~W; RIE power, 50~W; treatment time, 5~min)
in order to completely deplete surface-near electrons, resulting in
an upward surface band bending and complete removal the surface conductivity.\citep{bierwagen2011PLOX,berthold_plasma_2016}
During this treatment, a high density of reactive oxygen species attach
to the In\textsubscript{2}O\textsubscript{3} surface, removing electrons
from the In\textsubscript{2}O\textsubscript{3} to form negatively
charged adsorbates.\citep{berthold_plasma_2016} We found this adsorbate
layer to be stable against UV-illumination and to be removable only
by annealing the material.

The electrical sheet conductivity of the films under study is determined
by sheet resistance measurements in the commonly used van der Pauw
(vdP) arrangement. In combination with Hall effect measurements, which
directly provide the sheet electron concentration, this helps identify
the Hall electron mobility of the samples.

Since the measurements throughout this work are performed in ambient
environment, oxygen species from the air are expected to adsorb and
alter the transport properties of the SEAL. To circumvent this effect,
the samples under study have been exposed to UV illumination to force
desorption of those species. A light emitting diode (LED) that can
generate up to 12~mW ultraviolet (UV) A radiation with a wavelength
of 400~nm is utilized for this purpose. The corresponding photon
energy of 3.1\,eV is above the fundamental, dipole forbidden bandgap
and below the onset of strong optical absorption.\citep{Walsh2008}
The associated penetration depth in In$_{2}$O$_{3}$ is $\approx$1~\textgreek{m}m,\citep{Irmscher2013}
i.e., larger than the thickness of the investigated films. For most
measurements the LED is operated at a current of 13~mA, which corresponds
to approximately 8~mW of optical power, and the illuminated area
nominally covers the entire sample surface. This corresponds to a
photon flux of approximately $6\times10^{20}\,\mathrm{m^{-2}s^{-1}}$.
Due to the UV-induced desorption of species the conductivity of the
surface---and thus the total conductivity of the film---increases
with time until it starts saturating once a desorption--adsorption
equilibrium has been reached. Representative desorption/adsorption
cycles due to UV on/UV off periods can be found in Ref.~\onlinecite{rombach_SAB-2016}.
For the measurements to be reproducible, all samples are exposed to
UV for approximately 10~minutes, which has been found sufficient
to obtain desorption-adsorption equilibrium.

For the X-ray Photoelectron Spectroscopy (XPS) measurements, the samples
were mounted onto Ta sample holders, with the In\textsubscript{2}O\textsubscript{3}
layer electrically grounded, and inserted into an ultra-high vacuum
(UHV) system for surface analysis. The measurements were performed---after
preparation of the surface with UV illumination in vacuum at room
temperature---in normal emission using monochromated AlK\textgreek{a}
($h\nu=1486.7\,\mathrm{eV}$) radiation and a hemispherical electron
analyzer. More details about the setup and the experimental conditions
used for this study can be found in Ref.~\onlinecite{HIMMERLICH20076}.
The binding energy scale and the position of the Fermi level are regularly
calibrated for clean metal reference samples and the data analysis
was performed in analogy to the studies of UID and Mg-doped In\textsubscript{2}O\textsubscript{3}
films in Ref.~\onlinecite{berthold_plasma_2016}. The region around
the $E_{\mathrm{F}}$ was measured with an extended integration time.

Finally, the acquisition of the thermopower, otherwise known as Seebeck
coefficient, was performed as described in detail for In\textsubscript{2}O\textsubscript{3}
in Ref.~\onlinecite{preisslerPRB}. The Seebeck coefficient of the
SEAL has been calculated by the multilayer method described in Refs.~\onlinecite{Baron1969,PhysRevB.84.235302}
and then matched to the corresponding 2D electron concentrations---determined
by the Hall effect---for measurements with the same sheet resistance
(adjusted by proper UV illumination) using the van der Pauw method.
This is done because the Seebeck and Hall effect measurements are
performed in two separate systems and the sheet resistance is the
only property that can be measured in both setups and ensure same
surface conditions.

\section{Results and discussion}

\subsection{Transport properties of the adsorbate-attenuated and unattenuated
SEAL extracted by the multilayer method\label{subsec:SEAL_transport}}

\begin{figure}
\centering\includegraphics[width=8.5cm]{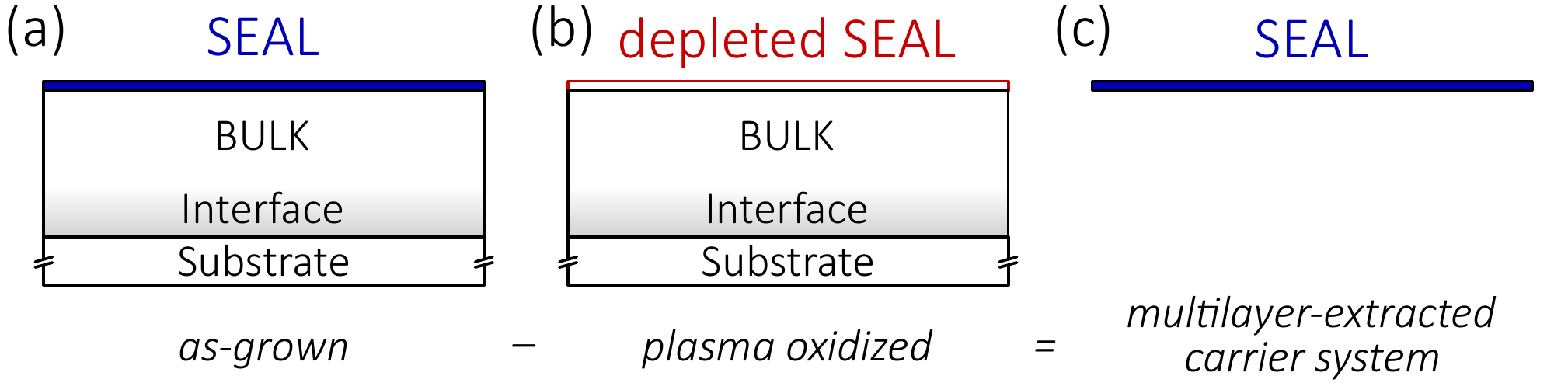}

\caption{Schematic representation of the carrier systems in the films under
study. The substrate is insulating, while it has been shown that the
samples under study possess a strong interface carrier system along
with the bulk of the film and the SEAL \citep{papadogianni_In2O3:Ni_YSZ}.
(a) All carrier systems included in the as-grown film, (b) depleted
SEAL in the plasma oxidized film, (c) extracted SEAL by the multilayer
method of Refs.~\onlinecite{Baron1969,PhysRevB.84.235302}. \label{fig:carrier_systems}}
\end{figure}

In order to extract the transport properties of the surface carrier
system of In\textsubscript{2}O\textsubscript{3} the multilayer method
described in Refs.~\onlinecite{Baron1969,PhysRevB.84.235302} will
be employed. Essentially, since all transport systems in our films
(depicted in Fig.~\ref{fig:carrier_systems}(a)) are connected in
parallel, the total sheet conductivity of the film will be the sum
of the separate sheet conductivities of the carrier systems comprising
it

\begin{equation}
G_{\mathrm{tot}}=G_{\mathrm{B}}+G_{\mathrm{I}}+G_{\mathrm{S}}\label{eq:G_total}
\end{equation}
where the subscripts indicate the \textit{bulk} (B), \textit{interface}
(I), and \textit{surface} (S) sheet conductivity. Let us assume the
case depicted in Fig.~\ref{fig:carrier_systems} with two films,
(a) and (b), comprising of the same carrier systems---bulk and interface---with
the exception of the SEAL, which is not present in film (b). Based
on the model described, the sheet conductivity of the carrier system
these films differ by, (c), could be extracted by subtracting the
total sheet conductivities of the two films. A technique to deplete
the SEAL is thus required for this method to be applied.

The plasma oxidation of the surface described in the experimental
part can provide samples with depleted SEAL. This indicates that the
sheet conductivity of a plasma oxidized sample equals $G_{\mathrm{PLOX}}=G_{\mathrm{B}}+G_{\mathrm{I}}$.
Combining this with the multilayer model of Refs.~\onlinecite{Baron1969,PhysRevB.84.235302}
allows one to extract not only the sheet conductivity, but also the
entire set of transport properties of the SEAL by performing Hall
effect measurements on an unintentionally doped (UID) sample of In\textsubscript{2}O\textsubscript{3}
before and after plasma treatment. As an example, one can extract
the sheet conductivity of the SEAL as 
\begin{equation}
G_{\mathrm{S}}^{\mathrm{w/\,ads.}}=G_{\mathrm{a.g.}}-G_{\mathrm{PLOX}}\label{eq:G_SEAL_with_ads}
\end{equation}
where a.g. is used to denote the untreated (besides oxygen annealing),
as-grown state of the film. This of course would correspond to an
upper estimate of the sheet conductivity of the SEAL with the effect
of present air adsorbates (superscript ``w/~ads.''). In our UID,
single crystalline film we found such a SEAL to feature a sheet conductivity
of $G_{\mathrm{S}}=3.80\times10^{-6}\,\mathrm{S}$, which is significantly
lower than that of the photoreduced SEAL ($G\approx3\times10^{-4}\,\mathrm{S}$)
in Ref.~\onlinecite{rombach_SAB-2016} and suggests that the oxygen
adsorbates from the air almost completely deplete the SEAL.

Since the plasma oxidation of the surface could potentially deplete
part of the bulk, the extracted sheet conductivity from Eq.~\ref{eq:G_SEAL_with_ads}
could contain the near surface bulk conductivity that got depleted.
In order to avoid these effects being reflected on the extracted SEAL
transport properties, one can extract the sheet conductivity of the
adsorbate-free (superscript ``w/o~ads.'') SEAL as follows

\begin{equation}
G_{\mathrm{S}}^{\mathrm{w/o\,ads.}}=\left(G_{\mathrm{a.g.}}^{\mathrm{\mathrm{UV}}}-G_{\mathrm{a.g.}}^{\mathrm{dark}}\right)-\left(G_{\mathrm{PLOX.}}^{\mathrm{\mathrm{UV}}}-G_{\mathrm{PLOX}}^{\mathrm{dark}}\right)\label{eq:G_SEAL}
\end{equation}

assuming the UV light exposure to remove all surface adsorbates by
photoreduction and full depletion of the SEAL in the dark by adsorbed
oxygen species. At this point, it should be pointed out that the high
penetration depth of the UV illumination could induce photoconduction
in the bulk of the material. Examination of the conductivity change
upon UV illumination of an undoped In\textsubscript{2}O\textsubscript{3}
film, whose surface had been depleted by undergoing the plasma oxidation
process, showed a sudden drop of the sheet conductivity by $2.82\times10^{-5}\,\mathrm{S}$,
which amounts up to 7\% of the total change in sheet conductivity
by the UV as observed in the untreated sample. This bulk photoconduction
effect is also excluded by the difference method of Eq.~\ref{eq:G_SEAL}.

The method described in Refs.~\onlinecite{Baron1969,PhysRevB.84.235302}
allows for the extraction of the full set of SEAL transport properties.
According to these, the mobility and Seebeck coefficient can both
be extracted in a similar manner using respectively 
\begin{multline}
\mu_{\mathrm{S}}^{\mathrm{w/o\,ads.}}=\frac{\mu_{\mathrm{a.g.}}^{\mathrm{UV}}\cdot G_{\mathrm{a.g.}}^{\mathrm{\mathrm{UV}}}-\mu_{\mathrm{a.g.}}^{\mathrm{dark}}\cdot G_{\mathrm{a.g.}}^{\mathrm{dark}}}{G_{\mathrm{S}}^{\mathrm{w/o\,ads.}}}\\
-\frac{\mu_{\mathrm{PLOX}}^{\mathrm{UV}}\cdot G_{\mathrm{PLOX.}}^{\mathrm{\mathrm{UV}}}-\mu_{\mathrm{PLOX}}^{\mathrm{dark}}\cdot G_{\mathrm{PLOX}}^{\mathrm{dark}}}{G_{\mathrm{S}}^{\mathrm{w/o\,ads.}}}\label{eq:mobility}
\end{multline}

\begin{multline}
S_{\mathrm{S}}^{\mathrm{w/o\,ads.}}=\frac{S_{\mathrm{a.g.}}^{\mathrm{UV}}\cdot G_{\mathrm{a.g.}}^{\mathrm{\mathrm{UV}}}-S_{\mathrm{a.g.}}^{\mathrm{dark}}\cdot G_{\mathrm{a.g.}}^{\mathrm{dark}}}{G_{\mathrm{S}}^{\mathrm{w/o\,ads.}}}\\
-\frac{S_{\mathrm{PLOX}}^{\mathrm{UV}}\cdot G_{\mathrm{PLOX.}}^{\mathrm{\mathrm{UV}}}-S_{\mathrm{PLOX}}^{\mathrm{dark}}\cdot G_{\mathrm{PLOX}}^{\mathrm{dark}}}{G_{\mathrm{S}}^{\mathrm{w/o\,ads.}}}\label{eq:seebeck}
\end{multline}

Finally, the sheet (2D) electron concentration of the SEAL without
the effect of air adsorbates can be easily calculated based on the
results of Eqs.~\ref{eq:G_SEAL} and \ref{eq:mobility} as

\begin{equation}
n_{\mathrm{S}}^{w/o\,ads}=\frac{G_{\mathrm{S}}^{\mathrm{w/o\,ads.}}}{q\mu_{\mathrm{n}}^{\mathrm{w/o\,ads.}}}\label{eq:electron_concentration}
\end{equation}

where $q$ is the elementary charge. Based on the equations above,
the SEAL of an undoped single-crystalline In\textsubscript{2}O\textsubscript{3}
film has been found to exhibit a sheet conductivity of $G_{\mathrm{S}}^{\mathrm{w/o\,ads.}}=3.26\times10^{-4}\,\mathrm{S}$,
a sheet electron concentration of $n_{\mathrm{S}}^{\mathrm{w/o\,ads.}}=1.45\times10^{13}\,\mathrm{cm^{-2}}$,
and a Hall electron mobility of $\mu_{\mathrm{S}}^{w/o\,ads}=155\,\mathrm{cm^{2}V^{-1}s^{-1}}$
without the effect of air adsorbates.

\subsection{Intentional attenuation by compensating acceptor doping}

\subsubsection*{Electrical transport}

\begin{figure}
\includegraphics[width=7cm]{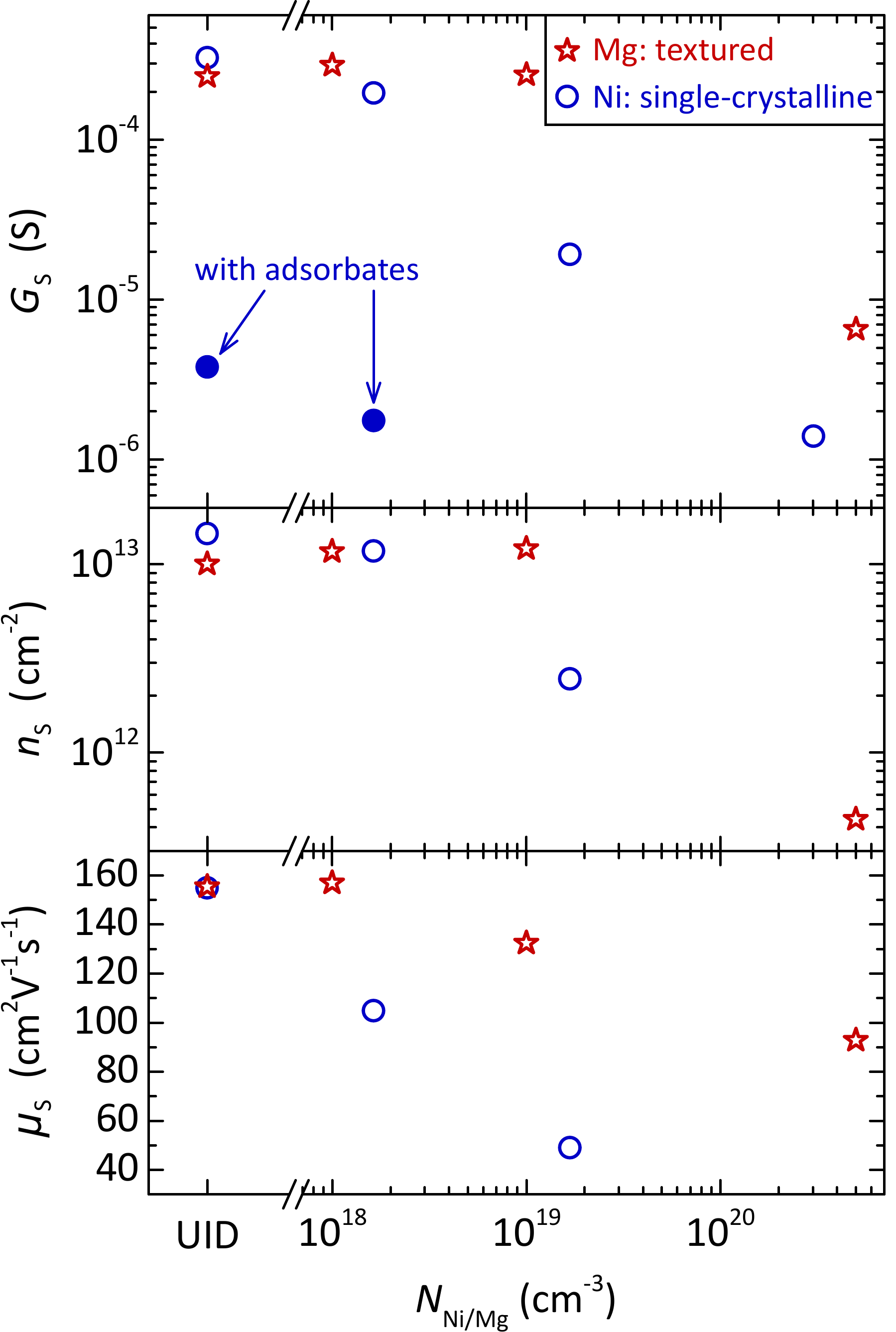}\centering

\caption{Transport properties of the SEAL as a function of compensating acceptor
concentration $N_{\mathrm{Ni/Mg}}$: Sheet conductivity $G_{\mathrm{S}}$,
sheet electron concentration $n_{\mathrm{S}}$, and Hall electron
mobility $\mu_{\mathrm{S}}$. The open symbols represent the extracted
data without the effect of air adsorbates (as in Eq.~\ref{eq:G_SEAL}.),
whereas the closed symbols correspond to the extracted SEAL transport
properties with air adsorbates (Eq.~\ref{eq:G_SEAL_with_ads}).\label{fig:SEAL_transport}}
\end{figure}

Figure~\ref{fig:SEAL_transport} depicts the SEAL transport properties
extracted from the Hall measurements by Eqs.\,(\ref{eq:G_SEAL},
\ref{eq:mobility}, and \ref{eq:electron_concentration}) of a series
of single-crystalline Ni-doped (blue circles) and textured Mg-doped
(red stars) films, along with their dedicated unintentionally doped
samples. Increasing compensating doping leads---as expected---to
a decrease in the extracted sheet conductivity of the SEAL for both
types of dopants and substrates. A Ni concentration of approximately
$2\times10^{19}\,\mathrm{cm^{-3}}$---which is comparable to the
peak surface electron concentration of Ref.~\onlinecite{King_2009_In2O3}---has
a significant effect on it, whereas a similar concentration of Mg,
$N_{\mathrm{Mg}}=10^{19}\,\mathrm{cm^{-3}}$ does not substantially
affect the SEAL transport properties. Higher Ni-doping $>10^{20}\,\mathrm{cm^{-3}}$
seems to deplete most of the surface carriers, reaching SEAL sheet
conductivities as low as $10^{-6}\,\mathrm{S}$ and a very low Hall
mobility, that does not allow for the extraction of a meaningful surface
electron concentration. Interestingly, an even higher Mg concentration
of $N_{\mathrm{Mg}}=5\times10^{20}\,\mathrm{cm^{-3}}$ does not fully
deplete the SEAL. Besides the doping ranges presented in Fig.~\ref{fig:SEAL_transport},
a higher Ni doped sample on YSZ (111) with $N_{\mathrm{Ni}}=2\times10^{21}\,\mathrm{cm^{-3}}$
has been also studied and shown (in Ref.~\onlinecite{papadogianni_In2O3:Ni_YSZ})
to be insulating, in which case all carrier systems---including the
SEAL---have been fully depleted.

To compare with the degree of depletion attained unintentionally by
air adsorbates, the data in Fig.~\ref{fig:SEAL_transport} represented
by full circles demonstrate the sheet conductivity of one UID and
one lightly Ni-doped sample that have been measured under dark conditions.
The effect of air adsorbates with an acceptorlike behavior is evidently
intense, as they decrease the sheet conductivity of the films by two
orders of magnitude.

Both the sheet electron concentration and mobility of the SEAL decrease
with increasing acceptor concentration, as would have been anticipated
for compensating dopants and the addition of charged scattering centers.

\textcolor{black}{However, there seems to be a different doping threshold
between the two sample series, which results in stronger or full depletion
of the SEAL, and they exhibit different mobilities. In particular,
the SEAL mobility of the single crystalline Ni-doped films is significantly
lower than the mobility of the Mg-doped ones on }\textsl{\textcolor{black}{c}}\textcolor{black}{-Al}\textsubscript{2}\textcolor{black}{O}\textsubscript{3}\textcolor{black}{,
which feature grain boundaries. This is a rather unexpected result,
however the lower electron concentration of the Ni-doped series could
possibly be attributed to the fact that YSZ is an oxygen conductor,
and therefore oxygen from the substrate could diffuse to the surface
and deplete the SEAL. Moreover, the position of the deep acceptors
in the band gap and, hence, the probability to compensate SEAL electrons,
can differ.}

\subsubsection*{X-ray photoelectron spectroscopy}

\begin{figure}
\includegraphics[width=6cm]{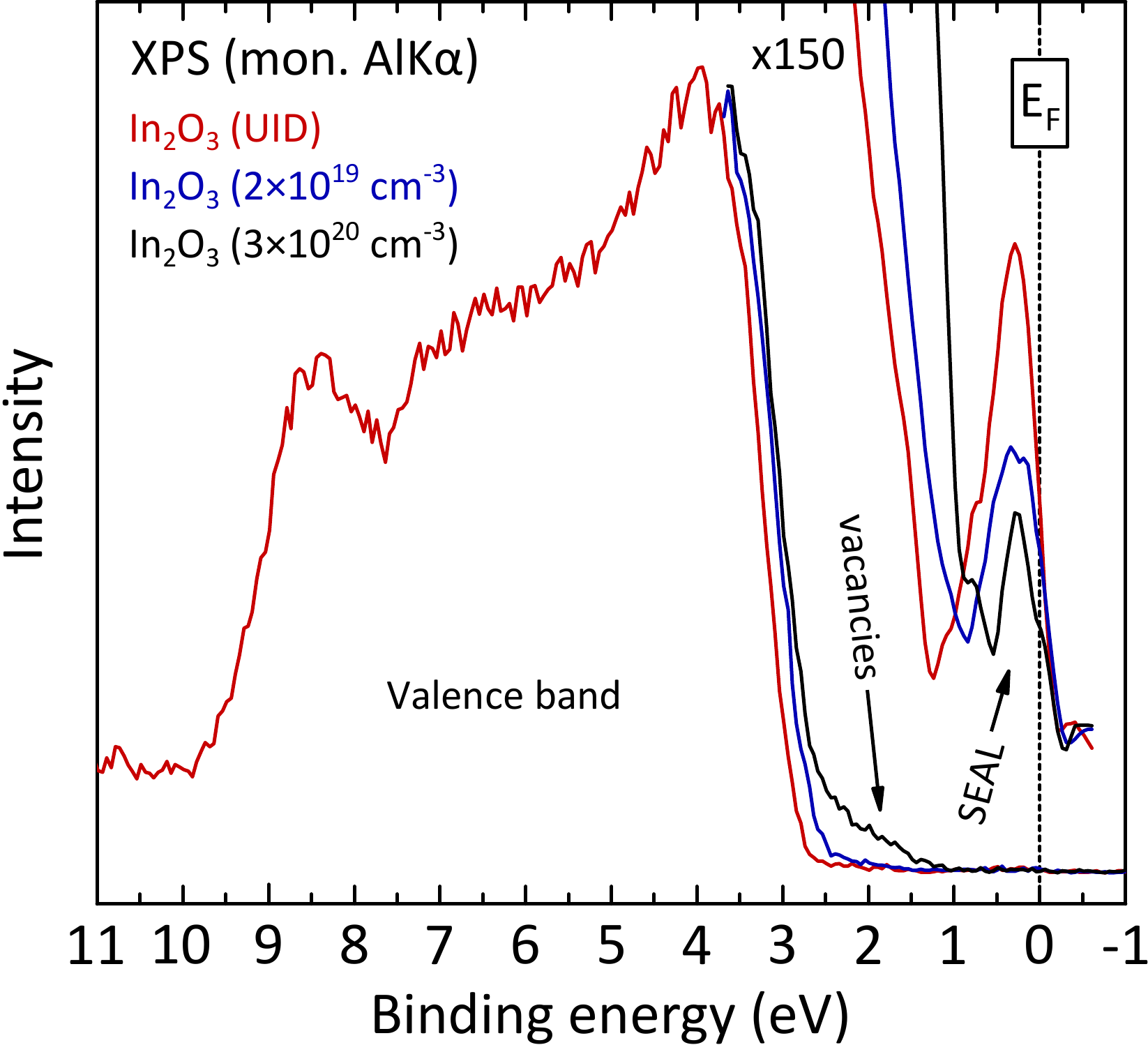}\centering

\caption{X-ray photoelectron spectra of the valence band and occupied conduction
band states of the single-crystalline unintentionally doped (UID)
and Ni-doped In\protect\textsubscript{2}O\protect\textsubscript{3}
films (Ni concentration as indicated). For the Ni-doped films, only
the VB edge is shown for clarity and to depict the occurring energy
shift and additional intragap states, as the shape of the complete
VB spectra at higher binding energies is similar to that of the UID
sample. The magnified region on the right has been smoothed (9-point
locally weighted smoothing).\label{fig:XPS_VB_spectrum}}
\end{figure}

\begin{figure}
\includegraphics[width=7cm]{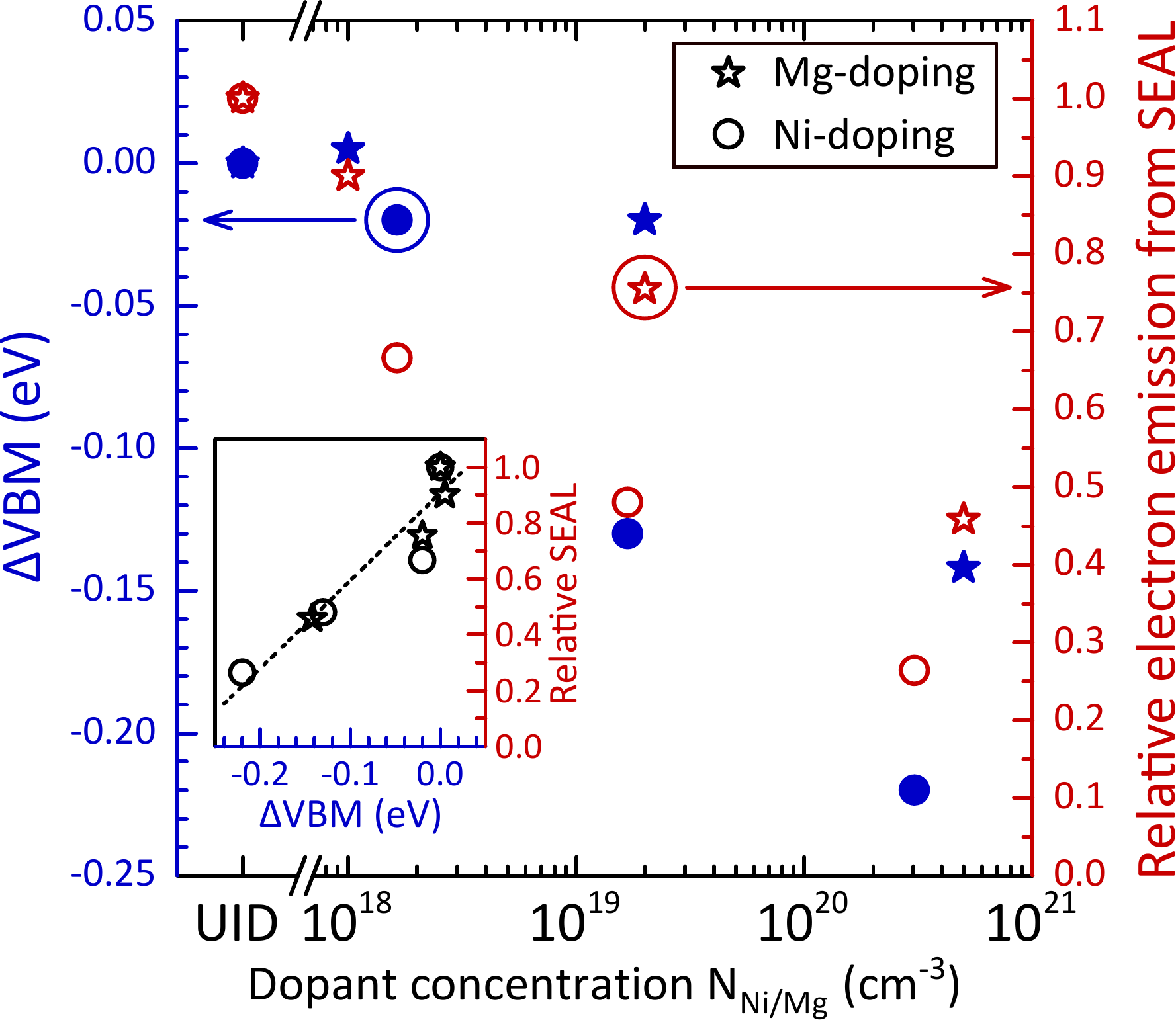}\centering

\caption{Change of valence band maximum \textgreek{D}VBM (blue, closed points)
and relative SEAL induced electron emission (red, open points) as
a function of acceptor concentration $N_{\mathrm{Ni/Mg}}$ extracted
from the XPS spectra. The relative electron emission values from the
SEAL have been normalized with respect to the unintentionally doped
reference sample. The inset shows the relative SEAL intensity (red
axis) vs. \textgreek{D}VBM (blue axis) including data from both In$_{2}$O$_{3}$:Ni
and In$_{2}$O$_{3}$:Mg.\label{fig:XPS_VBM_SEAL-area}}
\end{figure}

To relate the electrical transport results to the band structure and
surface band bending, XPS measurements have been performed on both
the Ni-doped single-crystalline and the Mg-doped textured samples.
The method probes the topmost few nanometers at the surface of the
film. Figure~\ref{fig:XPS_VB_spectrum} illustrates the valence band
spectrum obtained by XPS for the single-crystalline UID (red) and
the Ni-doped samples with $N_{\mathrm{Ni}}=2\times10^{19}\,\mathrm{cm^{-3}}$
and $3\times10^{20}\,\mathrm{cm^{-3}}$ (blue and black respectively),
after illumination with UV in vacuum, to best represent the state
of the films during the transport measurements. The binding energy
is presented with respect to the position of the Fermi level. The
broad distribution between 10 and 3~eV originates from emission of
valence band electrons\citep{Erhart_PhysRevB.75.153205,King_2009_In2O3,Himmerlich_JAP111_093704},
whereas the feature just below the $E_{\mathrm{F}}$ is due to partial
occupation of conduction band states of the SEAL \citep{king_SEAL-CNL-In2O3_2008,rombach_SAB-2016}
sustained by the distinct downward bending of the electronic bands
at the surface\citep{King_2009_In2O3}. When Ni acceptors are introduced
into the film, the valence band maximum (VBM) shifts towards lower
binding energy for higher Ni concentrations. The same effect is observed
for the core level energies. Both these shifts consistently indicate
a lowering of the surface Fermi level with increasing Ni concentration.
At the same time, the emission of electrons near the $E_{\mathrm{F}}$
from the SEAL is significantly lowered. The same effect is also observed
for the Mg-doped In\textsubscript{2}O\textsubscript{3} films. Furthermore,
for the highest Ni concentration, an enhanced emission is observed
above the VB maximum, which indicates the formation of intragap states.
Such states above the VBM in In\textsubscript{2}O\textsubscript{3}
have been attributed to the existence of oxygen vacancies\citep{Mizuno_1997,Tanaka_2002,Agoston_2009}.
Their appearance at high Ni concentrations could be an indication
that the incorporation of acceptors leads to a partial charge compensation
by formation of additional oxygen vacancies. However, this effect
has not been observed for highly Mg-doped In\textsubscript{2}O\textsubscript{3}
films\citep{berthold_plasma_2016}.

The determination of the absolute energy of the VBM and the difference
between surface $E_{\mathrm{F}}$ and conduction band minimum $E_{\mathrm{C_{SURF}}}-E_{\mathrm{F}}$
by XPS is not straightforward. We discuss our quantitative evaluation
in the supplemental material, which results in a value of $E_{\mathrm{C_{SURF}}}-E_{\mathrm{F}}=0.6\,\mathrm{eV}$
for UID In\textsubscript{2}O\textsubscript{3}. The change in band
edge position, however, can be determined by characterizing the energy
offset in XPS. We have compared the change of VBM (\textgreek{D}VBM)
for Ni- and Mg-doped films with varying acceptor concentrations. In
addition, the area of the emission near the $E_{\mathrm{F}}$ is used
as a quantitative measure of the electrons in the SEAL\citep{Berthold_PSSB_2018}.
Both values---\textgreek{D}VBM and the relative reduction of the
SEAL compared to UID films---are plotted as a function of the Ni
and Mg concentration in Fig.~\ref{fig:XPS_VBM_SEAL-area}. It is
evident that there is a correlation among the increasing acceptor
concentration, the shift of VB edge towards the $E_{\mathrm{F}}$,
and the reduction of surface electron concentration. Moreover, a roughly
linear relation between \textgreek{D}VBM and the relative electron
emission from SEAL is observed (inset of Fig.~\ref{fig:XPS_VBM_SEAL-area}).
In accordance with the electrical transport measurements, Ni-doping
is shown to induce stronger changes in the electronic properties in
comparison to the Mg-doping. Nonetheless, a significant depletion
of the SEAL has been achieved, i.e. down to 26\,\% of that of the
UID In\textsubscript{2}O\textsubscript{3} for the highest Ni concentration.
At this Ni concentration, however, transport measurements indicated
a stronger reduction of the SEAL (conductance decrease to less than
1\,\% of that of the UID In\textsubscript{2}O\textsubscript{3}).
We tentatively attribute this discrepancy to a residual O-adsorbate
coverage of the adsorption-desorption equilibrium during the transport
measurements under UV illumination in air.

\subsubsection*{Self-consistent Schrödinger--Poisson calculations of the near-surface
potential and electron density profiles}

\begin{table*}[t]
\caption{Schrödinger--Poisson calculations for various scenarios (for detailed
description see text). Values in blue have been calculated, while
the rest are fixed or predefined for each case.}
\begin{tabular}{ccccccccc}
\hline 
Nº  & $N_{\mathrm{D}}$  & $N_{\mathrm{A}}$  & $N_{\mathrm{A}}-N_{\mathrm{D}}$  & Acceptor type  & $E_{\mathrm{C_{BULK}}}-E_{\mathrm{F}}$  & $E_{\mathrm{C_{SURF}}}-E_{\mathrm{F}}$  & $N_{\mathrm{ss}}$  & $n_{\mathrm{S}}$\tabularnewline
 & (cm\textsuperscript{-3})  & (cm\textsuperscript{-3})  & (cm\textsuperscript{-3})  &  & (eV)  & (eV)  & (cm\textsuperscript{-2})  & (cm\textsuperscript{-2})\tabularnewline
\hline 
\hline 
\multicolumn{1}{c}{i} & \textcolor{black}{$3\times10^{17}$}  & \textcolor{black}{---}  &  & \textcolor{black}{---}  & \textcolor{blue}{0.047}  & \textcolor{black}{-0.600}  & \textcolor{blue}{$1.18\times10^{13}$(A)}  & \textcolor{blue}{$1.33\times10^{13}$(B)}\tabularnewline
ii  & \textcolor{black}{$3\times10^{17}$}  & \textcolor{blue}{$1.0\times10^{18}$}  & \textcolor{blue}{$7.0\times10^{17}$}  & mid-gap  & \textcolor{blue}{1.388}  & -0.380  & \textcolor{blue}{$6.84\times10^{12}$}  & \textcolor{black}{$3.46\times10^{12}$(0.26B)}\tabularnewline
iii  & \textcolor{black}{$3\times10^{17}$}  & \textcolor{blue}{$4.3\times10^{18}$}  & \textcolor{blue}{$4.0\times10^{18}$}  & mid-gap  & \textcolor{blue}{1.431}  & \textcolor{blue}{$-0.540$}  & \textcolor{black}{$1.18\times10^{13}$(A)}  & \textcolor{black}{$3.46\times10^{12}$(0.26B)}\tabularnewline
iv  & \textcolor{black}{$3\times10^{17}$}  & \textsl{\textcolor{blue}{$4.8\times10^{18}$}}  & \textcolor{blue}{$4.5\times10^{18}$}  & \textcolor{black}{shallow}  & \textcolor{blue}{2.762}  & \textcolor{black}{-0.380}  & \textcolor{black}{$1.18\times10^{13}$(A)}  & \textcolor{blue}{$7.70\times10^{9}$}\tabularnewline
v  & \textcolor{black}{$3\times10^{17}$}  & \textsl{\textcolor{blue}{$8.4\times10^{18}$}}  & \textcolor{blue}{$8.1\times10^{18}$}  & \textcolor{black}{mid-gap}  & \textcolor{blue}{1.449}  & \textcolor{black}{-0.380}  & \textcolor{black}{$1.18\times10^{13}$(A)}  & \textcolor{blue}{$1.17\times10^{10}$}\tabularnewline
\textcolor{black}{vi}  & \textbf{\textsl{\textcolor{blue}{$2.9546\times10^{20}$}}}  & \textcolor{black}{$3.0\times10^{20}$}  & \textcolor{blue}{$4.5\times10^{18}$}  & shallow  & \textcolor{blue}{2.762}  & \textcolor{black}{-0.380}  & \textcolor{black}{$1.18\times10^{13}$(A)}  & \textcolor{black}{$7.70\times10^{9}$}\tabularnewline
\textcolor{black}{vii}  & \textbf{\textsl{\textcolor{blue}{$2.9125\times10^{20}$}}}  & \textcolor{black}{$3.0\times10^{20}$}  & \textcolor{blue}{$8.7\times10^{18}$}  & \textcolor{black}{mid-gap}  & \textcolor{blue}{1.274}  & \textcolor{black}{-0.380}  & \textcolor{black}{$1.18\times10^{13}$(A)}  & \textcolor{black}{$1.17\times10^{10}$}\tabularnewline
\textcolor{black}{viii}  & \textbf{\textsl{\textcolor{blue}{$2.9961\times10^{20}$}}}  & \textcolor{black}{$3.0\times10^{20}$}  & \textcolor{blue}{$3.9\times10^{17}$}  & \textcolor{black}{shallow}  & \textcolor{blue}{2.697}  & \textcolor{black}{-0.380}  & \textcolor{blue}{$6.80\times10^{12}$}  & \textcolor{black}{$3.46\times10^{12}$(0.26B)}\tabularnewline
\textcolor{black}{ix}  & \textbf{\textsl{\textcolor{blue}{$2.9918\times10^{20}$}}}  & \textcolor{black}{$3.0\times10^{20}$}  & \textcolor{blue}{$8.2\times10^{17}$}  & \textcolor{black}{mid-gap}  & \textcolor{blue}{1.212}  & \textcolor{black}{-0.380}  & \textcolor{blue}{$6.84\times10^{12}$}  & \textcolor{black}{$3.46\times10^{12}$(0.26B)}\tabularnewline
x  & \textsl{\textcolor{blue}{$1.922\times10^{19}$}}  & \textcolor{black}{$2.0\times10^{19}$}  & \textcolor{blue}{$7.8\times10^{17}$}  & \textcolor{black}{mid-gap}  & \textcolor{blue}{1.281}  & \textcolor{black}{-0.380}  & \textcolor{blue}{$6.84\times10^{12}$}  & \textcolor{black}{$3.46\times10^{12}$(0.26B)}\tabularnewline
xi  & \textsl{\textcolor{blue}{$1.25\times10^{18}$}}  & \textcolor{black}{$2.0\times10^{18}$}  & \textcolor{blue}{$7.5\times10^{17}$}  & \textcolor{black}{mid-gap}  & \textcolor{blue}{1.351}  & \textcolor{black}{-0.380}  & \textcolor{blue}{$6.84\times10^{12}$}  & \textcolor{black}{$3.46\times10^{12}$(0.26B)}\tabularnewline
\hline 
\end{tabular}\label{SP_calc}\centering 
\end{table*}

\begin{figure}
\includegraphics[width=6cm]{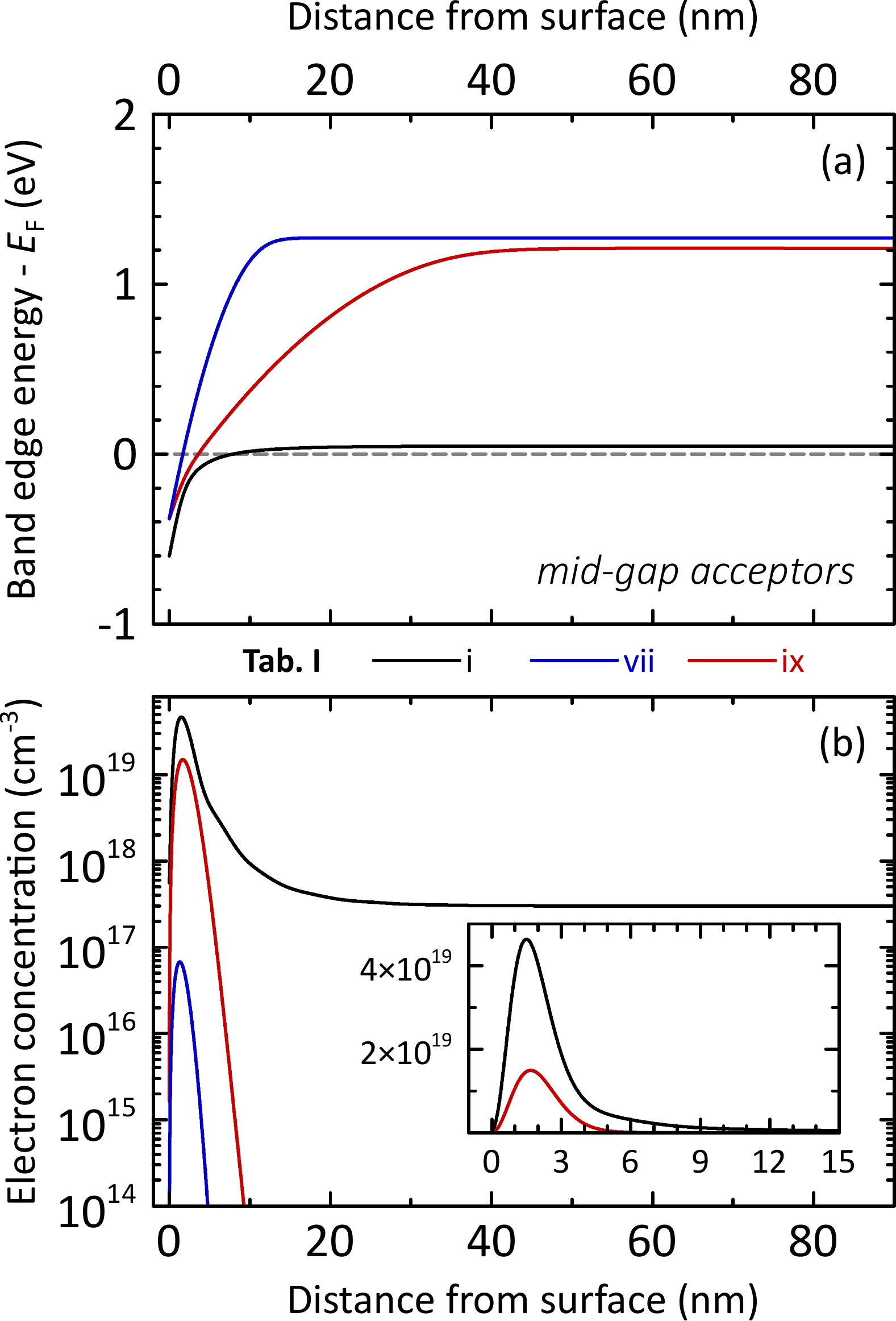}\caption{Band edge distributions (a) and electron concentration profiles (b)
of UID (i) and acceptor-doped In\protect\protect\protect\textsubscript{2}O\protect\protect\protect\textsubscript{3}
(vii, ix) calculated by Schrödinger--Poisson calculations considering
a deep acceptor level and constraints as shown in Tab.~\ref{SP_calc}.
The inset in (b) depicts two different $n_{\mathrm{S}}$ cases in
linear scale based on the experimental results from the XPS analysis
of the Ni-doped films. \label{fig:SP_depth-profiles}}
\end{figure}

To better understand the near-surface band and electron density profiles
in the In\textsubscript{2}O\textsubscript{3} films we have performed
self-consistent Schrödinger--Poisson calculations varying the acceptor
concentration $N_{\mathrm{A}}$ and using the experimentally determined
parameters obtained from electron transport measurements and X-ray
photoelectron spectroscopy as constraints. For the Poisson equation,
the Dirichlet boundary conditions (i.e., fixed surface potential relative
to the bulk Fermi level position) are used. We assume the full ionization
of donors present in the samples, whereas for acceptors the incomplete
ionization is also considered by defining the ionization energy level
relative to the valence band edge. The effect of the conduction band
nonparabolicity is also accounted for, and the band structure parameters
as well as the dielectric parameters have been taken from Ref.~\onlinecite{Feneberg_PhysRevB.93.045203}.
The net surface charge $N_{SS}$, corresponding to the difference
$(N_{D,S}^{+}-N_{A,S}^{-})$ of the 2D concentration of charged surface
donors and acceptors schematically shown in Fig.\,\ref{fig:doping-levels}(c),
is calculated from the charge neutrality condition applied to the
entire sample.

Table~\ref{SP_calc} and Figure~\ref{fig:SP_depth-profiles} summarize
the major parameters and profiles based on calculations under various
assumptions as described next.

Initially (Tab.~\ref{SP_calc}, case i) the $E_{\mathrm{C_{SURF}}}-E_{\mathrm{F}}$
from the XPS results corresponding to the UID sample, along with the
$E_{\mathrm{C_{BULK}}}-E_{\mathrm{F}}$ corresponding to a reasonable
bulk donor concentration of $N_{\mathrm{D}}=3\times10^{17}\,\mathrm{cm^{-3}}$
are used to calculate the SEAL concentration of the UID sample, $n_{\mathrm{S}}$.
The result of $n_{\mathrm{S}}=1.33\times10^{13}\,\mathrm{cm^{-2}}$
matches very well the experimentally extracted results from the Hall
effect measurements. $N_{\mathrm{D}}$ is assumed to correspond to
singly charged donors, ignoring the possible contribution from doubly
charged donors, like $V_{\mathrm{O}}^{2+}$ as previously discussed.

Next, acceptor doping is considered, where acceptors are not assumed
to induce the generation of compensating donors---i.e. the donor
concentration of $N_{\mathrm{D}}=3\times10^{17}\,\mathrm{cm^{-3}}$
is forced to be the same as for the unintentionally doped film in
case (i). If the $n_{\mathrm{S}}$ and $E_{\mathrm{C_{SURF}}}-E_{\mathrm{F}}$
from the XPS results for the sample doped with $N_{\mathrm{Ni}}=3\times10^{20}\,\mathrm{cm^{-3}}$
are considered (0.26B) (case ii), the result would be a reduced surface
states concentration, $N_{\mathrm{ss}}$, and an acceptor concentration,
$N_{\mathrm{A}}$, that is significantly lower than the actual doping
value. Alternatively, (case iii) fixing $N_{\mathrm{ss}}$ at the
value (A) of the UID sample, the predicted surface Fermi energy shift
of only $60\,\mathrm{meV}$ compared to the UID case does not match
the experimental results. If the $N_{\mathrm{ss}}$ is kept constant
at value (A) and the band edge shift from XPS are simultaneously considered
(case iv), the result would correspond to complete depletion of the
SEAL, even for significantly lower acceptor concentrations than the
actual $N_{\mathrm{Ni}}$. This result holds true irrespective of
the dopant position in the band gap, i.e. shallow (case iv) or mid-gap
(case v) acceptor level of Ni. However, as both the transport measurements
and the XPS results have revealed, the SEAL is still present for such
low acceptor concentrations. Thus, spontaneous generation of compensating
donors upon introduction of the acceptors has to be considered.

\textcolor{black}{Cases (vi) and (vii) show that even if the actual
acceptor concentration, }$N_{\mathrm{Ni}}$\textcolor{black}{, and
corresponding donor generation are considered, a fixed $N_{\mathrm{ss}}$
at the value (A) of the UID sample would predict the respective SEAL
to be fully depleted, for both, shallow or mid-gap acceptor levels.
Hence, the surface states concentration needs to be decreased in order
to reproduce the experimental results.}

Fixing the acceptor concentration at the intentional Ni-doping level
of $N_{\mathrm{Ni}}=3\times10^{20}\,\mathrm{cm^{-3}}$, a comparable
and, in fact, only slightly lower donor concentration (cases viii
and ix) has to be considered to result in corresponding SEAL concentration
$n_{\mathrm{S}}=(0.26B)=3.46\times10^{12}\,\mathrm{cm^{-2}}$ measured
by XPS. Once again, the position of the acceptor levels in the band
gap is of minimal significance to these results (viii, ix), with the
$N_{\mathrm{D}}$ and $N_{\mathrm{ss}}$ only being slightly affected.
Assuming deep acceptors---as there are indications that this is the
case\citep{Raebiger_PhysRevB.79.165202}, even though they are not
expected to be positioned in the middle of the band gap---and varying
their concentrations by one order of magnitude at each step (cases
ix-xi), we showcase that the relevant parameter for the reduction
of the SEAL is $N_{\mathrm{A}}-N_{\mathrm{D}}$. For shallow acceptors
the $N_{\mathrm{A}}-N_{\mathrm{D}}$ does not change at all (see supporting
information).

Even though these calculations rely on certain assumptions and constraints,
they allow to deduce important trends. The most important band edge
and electron concentration profiles of Tab.~\ref{SP_calc} are plotted
in Fig.~\ref{fig:SP_depth-profiles}. The inset shows the carrier
distribution in a linear scale for three different $n_{\mathrm{S}}$
cases, matching the relative emission of the SEAL derived from XPS,
and highlights the reduction of surface electron concentration when
Ni or Mg acceptors are incorporated into the In\textsubscript{2}O\textsubscript{3}
layers. \textcolor{black}{Integration of the profile in the inset
of Fig.~\ref{fig:SP_depth-profiles} corresponding to the acceptor-doped
case (ix) indicates that the majority (approximately 90\%) of the
SEAL carriers are lying within 3~nm from the surface.}

Consequently, no complete depletion of the In\textsubscript{2}O\textsubscript{3}
SEAL is achieved by acceptors at these doping concentrations, due
to the spontaneous formation of compensating donors. This conclusion
contrasts the findings in Mg-doped InN\citep{linhart_Mg-dop_InN_SEAL},
where the lack of compensating donors results in the immense reduction
of the SEAL. Nevertheless, the obtained results are promising towards
the tunability of the In\textsubscript{2}O\textsubscript{3} surface
properties and the expansion of its potential device applications.

\subsection{Enhanced thermopower in the SEAL 2DEG}

\begin{figure}
\includegraphics[width=7cm]{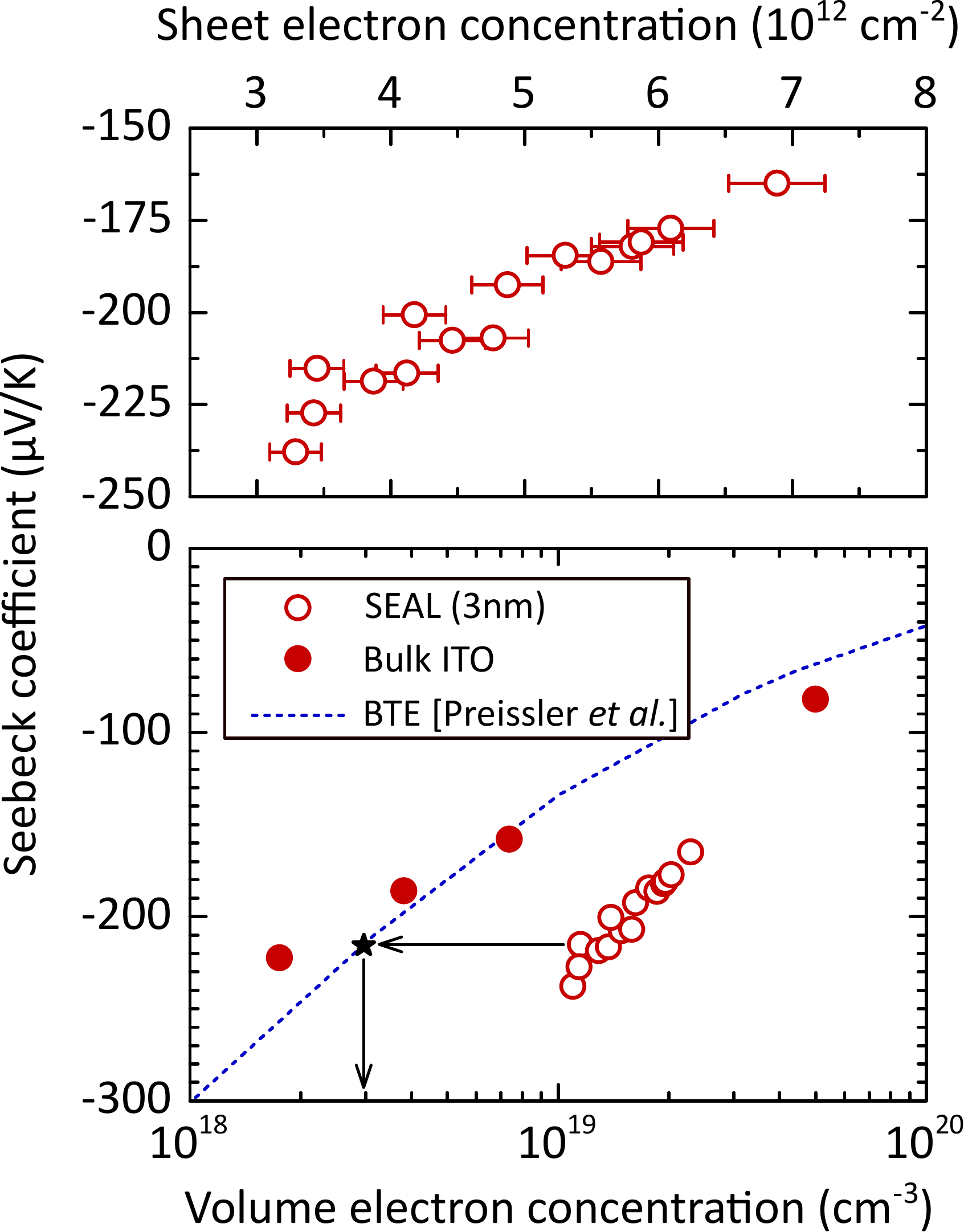}\centering

\caption{Top: Seebeck coefficient as a function of the surface sheet electron
concentration measured by Hall effect experiments in the vdP arrangement.
Bottom: comparison of the volume electron concentration of the SEAL
with that of bulk ITO films and solution of the Boltzmann transport
equation for $m^{*}=0.3m_{\mathrm{e}}$ from the work of \citet{preisslerPRB}
(Fig.~9 therein).\label{fig:Seebeck_SEAL-vs-ITO}}
\end{figure}

For the determination of the Seebeck coefficient of the In\textsubscript{2}O\textsubscript{3}
surface, the Mg-doped sample with $N_{\mathrm{Mg}}=10^{19}\,\mathrm{cm^{-3}}$---with
an intact SEAL and almost fully depleted parallel carrier systems---has
been utilized. The effect of a weak parallel interface system has
been taken into consideration and excluded using the multilayer method
described previously, and specifically Eq.~\ref{eq:seebeck}. Since
the oxygen plasma treatment did not essentially affect the transport
properties of this particular sample, the Seebeck coefficient was
ultimately extracted from the measurements of the dark and photoreduced
states of the as-grown film. The sheet electron concentration of the
films has been gradually modulated using UV illumination with varying
optical power between 1 and 12~mW and corresponding waiting time,
which results in adsorbed, acceptorlike, oxygen species between high
and low steady-state coverage. As shown in Fig.~\ref{fig:Seebeck_SEAL-vs-ITO}
(top), the Seebeck coefficient of the SEAL is negative, as expected
for carrier systems with their majority carriers being electrons,
and exhibits a decreasing magnitude with increasing electron concentration.
Due to limitations of the external UV illumination at the Seebeck
setup, it has not been possible to obtain a state of the In\textsubscript{2}O\textsubscript{3}
SEAL with higher electron concentration.

According to Ref.~\onlinecite{Shimizu6438}, the Seebeck coefficient
of a two-dimensional electron gas (2DEG)---in that case ZnO---exhibits
an increased absolute value, in comparison to that of a three-dimensional
electron gas (3DEG), if the semiconductor is well degenerate. This
effect is also displayed in Fig.~\ref{fig:Seebeck_SEAL-vs-ITO},
where the Seebeck coefficient of the SEAL (2DEG) is compared to the
experimental values of four ITO films (3DEG) as well as the theoretical
curve of Ref.~\onlinecite{preisslerPRB} based on the solution of
the Boltzmann transport equation (BTE) for similar volume electron
concentrations as that of the SEAL. This comparison yields a $\approx80$\,\%
larger Seebeck coefficient of the surface 2DEG compared to a bulk
3DEG with comparable volume electron concentration. The average volume
electron concentration of the SEAL has been calculated assuming that
the vast majority of the carriers lie within a 3~nm thick system
(cf. SP calculations of case (ix) in Tab.\,\ref{SP_calc} and Fig.\,\ref{fig:SP_depth-profiles}
for similar sheet electron concentration). That of the homogeneously
Sn-doped ITO films has been obtained from Hall effect measurements
and the film thickness.

Approaching it from a different perspective, if the Seebeck coefficient
of the SEAL would coincide with the BTE solution of a bulk system
in Fig.~\ref{fig:Seebeck_SEAL-vs-ITO}, as indicated by the black
arrow for one of the data points, the corresponding average volume
electron concentration of the SEAL would be significantly lower. As
explained in detail in Ref.~\onlinecite{papadogianni_APL_2015},
the sheet and volume electron concentrations of a carrier system can
be used to estimate its effective thickness, $t_{\mathrm{eff}}=\frac{n_{\mathrm{2D}}}{n_{\mathrm{3D}}}$.
If the volume electron concentration dictated by the BTE curve, which
is valid for bulk systems, is combined with the sheet electron concentration
directly measured by Hall, the resulting SEAL thicknesses would vary
between $11-15\,\mathrm{nm}$. This is a clear overestimation in comparison
to both the findings of the SP calculations in Fig.~\ref{fig:SP_depth-profiles}
and previous works\citep{king_SEAL-CNL-In2O3_2008,Nagata_2019,Nagata_PES_In2O3},
again indicating an enhanced thermopower of the SEAL.

\section{Summary and Conclusion}

In this work we have experimentally determined the transport properties
of the In\textsubscript{2}O\textsubscript{3} surface electron accumulation
layer by Hall effect measurements through applying a dual-layer model
in combination with plasma oxidation treatments of the surface. Oxygen
adsorbates from the ambient air almost completely deplete the SEAL
of an unintentionally doped film reducing its sheet conductivity to
$G_{\mathrm{S}}=3.8\times10^{-6}\,\mathrm{S}$. Illuminating the surface
with UV radiation largely removed the oxygen adsorbates, resulting
in a SEAL with a sheet conductivity of $G_{\mathrm{S}}=3.26\times10^{-4}\,\mathrm{S}$,
a sheet electron concentration of $n_{\mathrm{S}}=1.45\times10^{13}\,\mathrm{cm^{-2}}$,
and a Hall electron mobility of $\mu_{\mathrm{S}}=155\,\mathrm{cm^{2}V^{-1}s^{-1}}$.
We further demonstrated a gradual reduction of this SEAL by increasing
compensating bulk acceptor doping with two different elements, namely
Ni and Mg, and achieved nearly complete depletion with $N_{\mathrm{Ni}}=3\times10^{20}\,\mathrm{cm^{-3}}$
doping. The gradual depletion of the SEAL with doping concentration
has been confirmed by XPS measurements, able to determine the position
of the valence band maximum and SEAL peak area close to the Fermi
level. These results were further supported by Schrödinger--Poisson
calculations, which clearly show that the introduction of acceptors
in the In\textsubscript{2}O\textsubscript{3} results in the subsequent
generation of comparable concentrations of compensating donors. This
result holds true irrespective of the position of the acceptors in
the band gap, i.e. whether the corresponding levels are deep (mid-gap)
or shallow. This mechanism hinders the complete depletion of the In\textsubscript{2}O\textsubscript{3}
SEAL. However, our results showing significant attenuation of the
surface with acceptor doping are still valuable for device applications
requiring tunable surface transport properties. Ultimately, the thermopower
of the In\textsubscript{2}O\textsubscript{3} SEAL is investigated.
In agreement with previous studies on ZnO, the Seebeck coefficient
of the 2DEG at the In\textsubscript{2}O\textsubscript{3} surface
is shown to be enhanced by $\approx80$\,\% in comparison to the
3DEG with comparable volume electron concentration in bulk ITO films.

\section*{Acknowledgment}

We would like to thank Y. Takagaki for critically reading this manuscript,
F. Gutsche for technical support with the Seebeck setup, and W. Anders
for the oxygen plasma treatment of the samples. This study was performed
in the framework of GraFOx, a Leibniz ScienceCampus partially funded
by the Leibniz association. We are grateful for the financial support
by the Deutsche Forschungsgemeinschaft (grants BI 1754/1-1 and HI
1800/1-1).

 \bibliographystyle{apsrev4-1}
\bibliography{SEAL_In2O3_literature}

\end{document}

\title{SUPPLEMENTAL MATERIAL\\
The two-dimensional electron gas of the In\textsubscript{2}O\textsubscript{3}
surface: Enhanced thermopower, electrical transport properties, and
its reduction by adsorbates or compensating acceptor doping}
\author{Alexandra Papadogianni}
\affiliation{Paul-Drude-Institut für Festkörperelektronik, Leibniz-Institut im
Forschungsverbund Berlin e.V., Hausvogteiplatz 5\textendash 7, D\textendash 10117
Berlin, Germany}
\author{Julius Rombach}
\affiliation{Paul-Drude-Institut für Festkörperelektronik, Leibniz-Institut im
Forschungsverbund Berlin e.V., Hausvogteiplatz 5\textendash 7, D\textendash 10117
Berlin, Germany}
\author{Theresa Berthold}
\affiliation{Institut für Physik und Institut für Mikro- und Nanotechnologien,
Technische Universität Ilmenau, PF 100565, D\textendash 98684 Ilmenau,
Germany}
\author{Vladimir Polyakov}
\affiliation{Fraunhofer-Institut für Angewandte Festkörperphysik, Tullastraße 72,
D-79108 Freiburg, Germany}
\author{Stefan Krischok}
\affiliation{Institut für Physik und Institut für Mikro- und Nanotechnologien,
Technische Universität Ilmenau, PF 100565, D\textendash 98684 Ilmenau,
Germany}
\author{Marcel Himmerlich}
\affiliation{Institut für Physik und Institut für Mikro- und Nanotechnologien,
Technische Universität Ilmenau, PF 100565, D\textendash 98684 Ilmenau,
Germany}
\affiliation{CERN, European Organization for Nuclear Research, 1211 Meyrin, Switzerland}
\author{Oliver Bierwagen}
\affiliation{Paul-Drude-Institut für Festkörperelektronik, Leibniz-Institut im
Forschungsverbund Berlin e.V., Hausvogteiplatz 5\textendash 7, D\textendash 10117
Berlin, Germany}
\maketitle

\subsection*{\textit{Extrapolation of the valence band edge from XPS}}

\begin{figure}
\includegraphics[width=8cm]{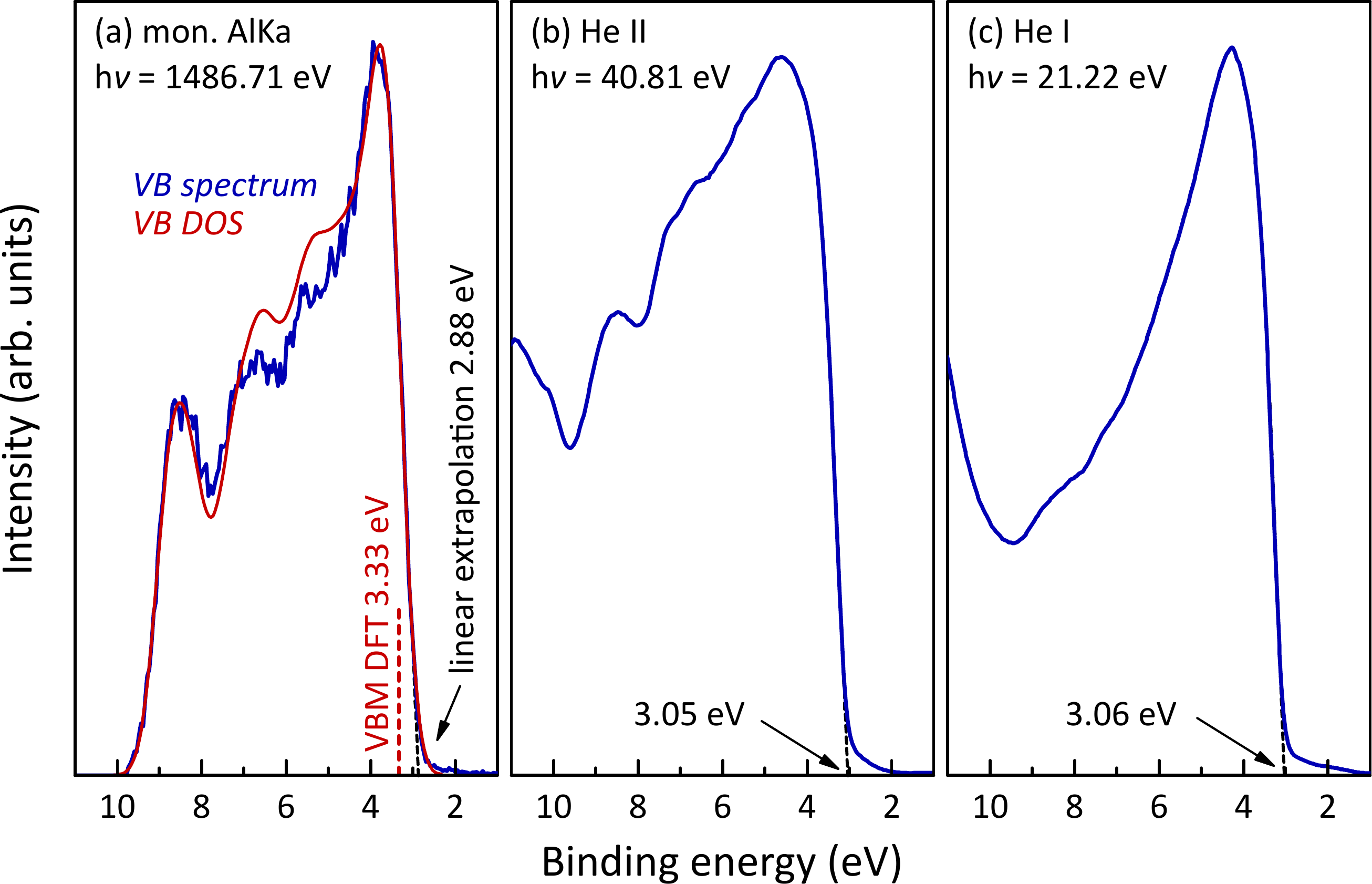}\centering\label{VB_In2O3}

\caption{Comparison of the VB photoelectron spectra of a vacuum-annealed In\protect\textsubscript{2}O\protect\textsubscript{3}
reference sample excited by different light sources\textemdash from
left to right: (a) AlK\textgreek{a}, (b) He II and (c) He I radiation.
The given numbers are the energy at the intersect of the linear extrapolation
of the VB edge with the abscissa. The red spectrum in (a) represents
the broadened VB density of states (DOS) of In\protect\textsubscript{2}O\protect\textsubscript{3}
from density-functional theory. The DOS has been shifted to match
the experimental data and the corresponding VBM of the non-boradened
DOS at $T=0\,\mathrm{K}$ is $3.33\,\mathrm{eV}$ below the Fermi
energy.}
\end{figure}

The determination of the absolute position of the surface valence
band maximum (VBM) with respect to the Fermi energy ($E_{\mathrm{F}}$)
based on X-ray photoelectron spectroscopy (XPS) measurements is not
straightforward, due to the broadening of the spectral features induced
by the relative high line width (the typical Fermi edge width is approximately
300~meV). Figure~\ref{VB_In2O3}~(a) shows an XPS valence band
spectrum of an uncontaminated In\textsubscript{2}O\textsubscript{3}
surface. The linear extrapolation of the VB edge gives a value of
2.88~eV, while the comparison with the broadened VB density of states\citep{King_2009_In2O3}
at $T=0\,\mathrm{K}$ reveals a VBM position of 3.33~eV. We consider
that this value would overestimate the actual VBM, since a thermal
broadening as well as uncertainties in DFT calculations have to be
considered. Comparing the corresponding ultraviolet photoelectron
spectroscopy (UPS) VB spectra with higher surface sensitivity, we
obtain a VBM of approximately 3.1~eV below the $E_{\mathrm{F}}$.
In this case, the experimental resolution is in the range of 50~meV,
which is considerably low to allow for reliable estimations. Considering
the band gap range of In\textsubscript{2}O\textsubscript{3} (2.6\textendash 2.8~eV)
and the possible band gap renormalization, due to conduction band
occupation below 100~eV, in the case where the electron concentration
stays in the $10^{19}\,\mathrm{cm^{-3}}$ range\citep{Feneberg_PhysRevB.93.045203},
we estimate the position of the conduction band minimum for the UID
In\textsubscript{2}O\textsubscript{3} to be 0.6~eV below the $E_{\mathrm{F}}$.

For samples that have been exposed to air after growth, UPS measurements
are not applicable to determine the VBM, since the spectral features
mainly include features of the surface adsorbates. We did not intend
to use thermal vacuum treatments to clean the surface of the Ni- and
Mg-doped films from adsorbates, which could potentially induce modifications
of the electronic properties, doping characteristics and possibly
induce other types of defects (mainly creation of donors due to oxygen
vacancy formation). Consequently, such UPS measurements could not
be implemented in this case. However, having defined a reference value
for the XPS VB spectrum of clean UID In\textsubscript{2}O\textsubscript{3}
as described above, we can rely on the VB shifts determined by extrapolation
of the XPS spectra, since changes in band bending affect the binding
energy scale directly. It is often argued that XPS averages over a
certain depth range and hence does not directly allow to determine
VB shifts correctly. It needs to be pointed out that the main signal
still originates directly from the topmost few nanometers of the surface
and a linear extrapolation determines the upper edge of the VB depth
profile, even though the high binding energy side might be broadened
or narrowed. The trends of band edge shifts are anyways reflected
correctly and there might only exists a slight underestimation of
\textgreek{D}VBM for the presented measurements.

\subsection*{\textit{Detailed Schrödinger\textendash Poisson calculations and
the case of shallow acceptors}}

\begin{figure*}[!]
\centering\includegraphics[width=15cm]{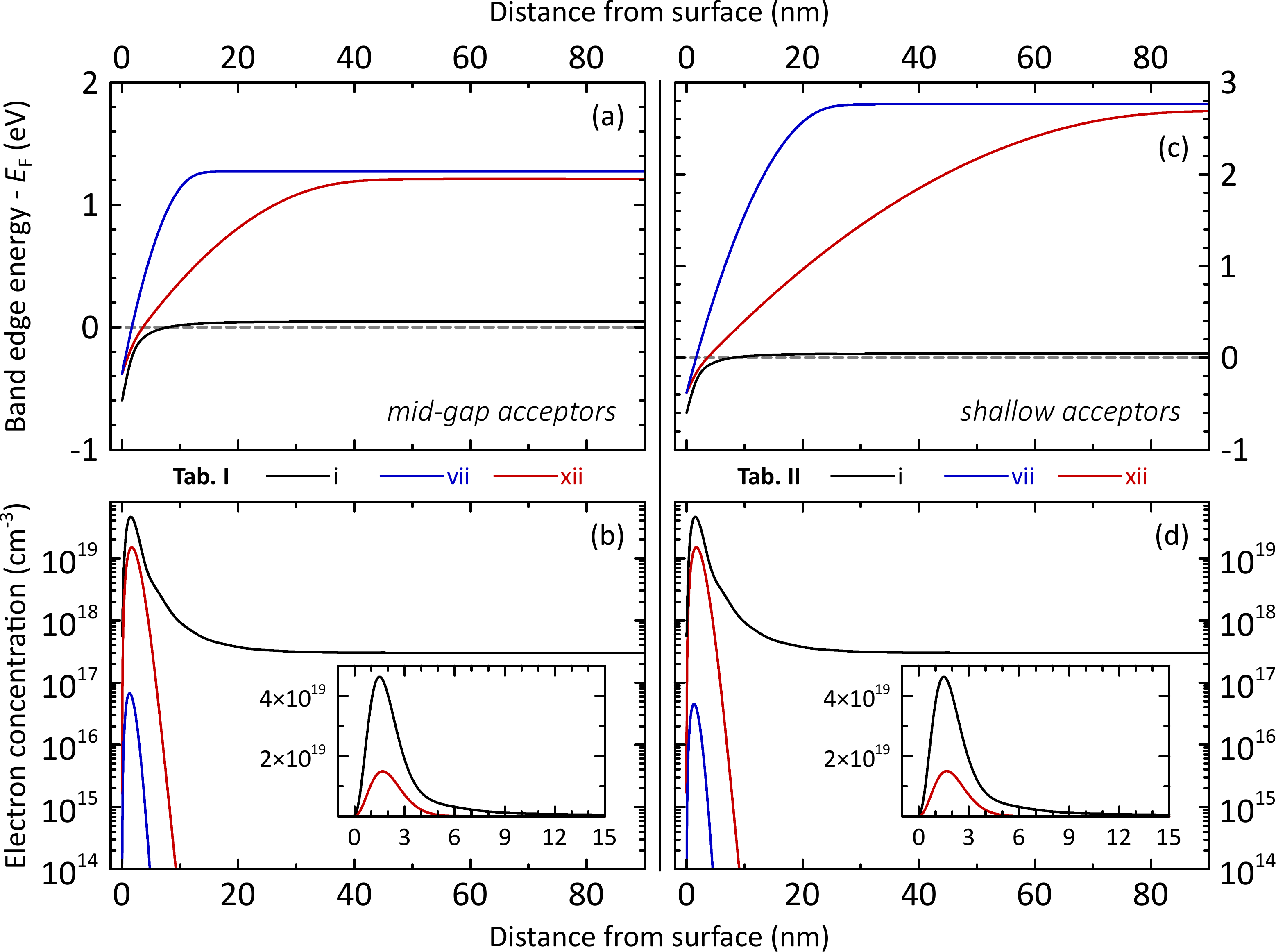}

\caption{Band edge distributions (a, c) and electron concentration profiles
(b, d) of UID (i) and acceptor-doped In\protect\textsubscript{2}O\protect\textsubscript{3}
(vii, xii) calculated by Schrödinger\textendash Poisson calculations
considering either deep (a, b) or shallow acceptor levels (c, d) and
constraints as shown in Tab.~\ref{SP-deep} and Tab.~\ref{SP-shallow}
respectively. The insets depict different $n_{\mathrm{S}}$ cases
in linear scale based on the experimental results from the XPS analysis
of the Ni-doped films. \label{fig:SP_depth-profiles-SUPP}}
\end{figure*}

Tables~\ref{SP-deep} and \ref{SP-shallow} show the full list of
Schrödinger\textendash Poisson (SP) calculations performed to get
a deeper understanding of the band alignment and transport properties
of the SEAL in acceptor-doped In\textsubscript{2}O\textsubscript{3}
for the case of deep (mid-gap)\textendash extensively discussed in
the main manuscript\textemdash and shallow acceptor levels respectively.
These tables also include the cases discussed in detail in the main
manuscript. Figure~\ref{fig:SP_depth-profiles-SUPP} presents the
depth profiles of the band edge energy relative to the Fermi level
$E_{\mathrm{F}}$ and electron concentration for mid-gap (left, exactly
as presented in the manuscript) and shallow (right) acceptors. The
profiles describe the UID (i) and acceptor-doped cases matching the
$n_{\mathrm{S}}$ observed by XPS for the Ni-doped sample with $N_{\mathrm{Ni}}=3\times10^{20}\,\mathrm{cm^{-3}}$
(vii) and assuming that the concentration of surface states remains
unchanged (xii), which proves to incorrectly represent the real experimental
results, as it leads to a depleted SEAL. Comparing the deep and shallow
acceptor cases one can conclude that the position of the acceptor
level in the band gap does not significantly affect the SEAL. The
parameter $N_{\mathrm{A}}-N_{\mathrm{D}}$ is the most relevant to
obtain a certain value for $n_{\mathrm{S}}$, if a fixed value for
$E_{\mathrm{C_{SURF}}}-E_{\mathrm{F}}$ is taken into consideration.
This becomes particularly apparent for the shallow acceptor case,
where the $N_{\mathrm{A}}-N_{\mathrm{D}}$ remains constant: see Tab.~\ref{SP-shallow}
realistic cases (iii-vii) describing a $n_{\mathrm{S}}=3.46\times10^{12}\,\mathrm{cm^{-2}}$
and cases (ix-xii) assuming an unchanged $N_{\mathrm{ss}}$ and depleted
SEAL.

\begin{table*}
\caption{Schrödinger\textendash Poisson calculations for various scenarios
assuming \uline{deep} (mid-gap) acceptor levels lying close to
the middle of the band gap. Values in blue have been calculated, while
the rest are fixed or predefined for each case.}
\begin{tabular}{cccccccc}
\hline 
Case & $N_{\mathrm{D}}$ & $N_{\mathrm{A}}$ & $N_{\mathrm{A}}-N_{\mathrm{D}}$ & $E_{\mathrm{C_{BULK}}}-E_{\mathrm{F}}$ & $E_{\mathrm{C_{SURF}}}-E_{\mathrm{F}}$ & $N_{\mathrm{ss}}$ & $n_{\mathrm{S}}$\tabularnewline
 & (cm\textsuperscript{-3}) & (cm\textsuperscript{-3}) & (cm\textsuperscript{-3}) & (eV) & (eV) & (cm\textsuperscript{-2}) & (cm\textsuperscript{-2})\tabularnewline
\hline 
i & \textcolor{black}{$3\times10^{17}$} & \textcolor{black}{\textemdash{}} &  & \textcolor{blue}{0.047} & -0.600 & \textsl{\textcolor{blue}{$1.18\times10^{13}$}}\textcolor{blue}{(A)} & \textsl{\textcolor{blue}{$1.33\times10^{13}$}}\textcolor{blue}{(B)}\tabularnewline
ii & \textcolor{black}{$3\times10^{17}$} & \textsl{\textcolor{blue}{$6.7\times10^{17}$}} & \textsl{\textcolor{blue}{$3.7\times10^{17}$}} & \textcolor{blue}{1.370} & -0.470 & \textsl{\textcolor{blue}{$8.75\times10^{12}$}} & \textsl{\textcolor{black}{$6.38\times10^{12}$}}\textcolor{black}{(0.48B)}\tabularnewline
iii & \textcolor{black}{$3\times10^{17}$} & \textsl{\textcolor{blue}{$1.0\times10^{18}$}} & \textsl{\textcolor{blue}{$7.0\times10^{17}$}} & \textcolor{blue}{1.388} & -0.380 & \textsl{\textcolor{blue}{$6.84\times10^{12}$}} & \textsl{\textcolor{black}{$3.46\times10^{12}$}}\textcolor{black}{(0.26B)}\tabularnewline
iv & \textsl{\textcolor{blue}{$1.25\times10^{18}$}} & \textcolor{black}{$2.0\times10^{18}$} & \textsl{\textcolor{blue}{$7.5\times10^{17}$}} & \textcolor{blue}{1.351} & -0.380 & \textsl{\textcolor{blue}{$6.84\times10^{12}$}} & \textsl{\textcolor{black}{$3.46\times10^{12}$}}\textcolor{black}{(0.26B)}\tabularnewline
v & \textsl{\textcolor{blue}{$1.921\times10^{19}$}} & \textcolor{black}{$2.0\times10^{19}$} & \textsl{\textcolor{blue}{$7.9\times10^{17}$}} & \textcolor{blue}{1.281} & -0.380 & \textsl{\textcolor{blue}{$6.84\times10^{12}$}} & \textsl{\textcolor{black}{$3.46\times10^{12}$}}\textcolor{black}{(0.26B)}\tabularnewline
vi & \textsl{\textcolor{blue}{$1.9918\times10^{20}$}} & \textcolor{black}{$2.0\times10^{20}$} & \textsl{\textcolor{blue}{$8.2\times10^{17}$}} & \textcolor{blue}{1.222} & -0.380 & \textsl{\textcolor{blue}{$6.84\times10^{12}$}} & \textsl{\textcolor{black}{$3.46\times10^{12}$}}\textcolor{black}{(0.26B)}\tabularnewline
vii & \textsl{\textcolor{blue}{$2.9918\times10^{20}$}} & \textcolor{black}{$3.0\times10^{20}$} & \textsl{\textcolor{blue}{$8.2\times10^{17}$}} & \textcolor{blue}{1.212} & -0.380 & \textsl{\textcolor{blue}{$6.84\times10^{12}$}} & \textsl{\textcolor{black}{$3.46\times10^{12}$}}\textcolor{black}{(0.26B)}\tabularnewline
viii & \textcolor{black}{$3\times10^{17}$} & \textsl{\textcolor{blue}{$4.3\times10^{18}$}} & \textsl{\textcolor{blue}{$4.0\times10^{18}$}} & \textcolor{blue}{1.431} & \textcolor{blue}{-0.540} & \textcolor{black}{$1.18\times10^{13}$(A)} & \textsl{\textcolor{black}{$3.46\times10^{12}$}}\textcolor{black}{(0.26B)}\tabularnewline
ix & \textcolor{black}{$3\times10^{17}$} & \textsl{\textcolor{blue}{$8.4\times10^{18}$}} & \textsl{\textcolor{blue}{$8.1\times10^{18}$}} & \textcolor{blue}{1.449} & -0.380 & \textcolor{black}{$1.18\times10^{13}$(A)} & \textsl{\textcolor{blue}{$1.17\times10^{10}$}}\tabularnewline
x & \textsl{\textcolor{blue}{$1.1692\times10^{19}$}} & \textcolor{black}{$2.0\times10^{19}$} & \textsl{\textcolor{blue}{$8.3\times10^{18}$}} & \textcolor{blue}{1.355} & -0.380 & \textcolor{black}{$1.18\times10^{13}$(A)} & \textsl{\textcolor{black}{$1.17\times10^{10}$}}\tabularnewline
xi & \textsl{\textcolor{blue}{$1.9132\times10^{20}$}} & \textcolor{black}{$2.0\times10^{20}$} & \textsl{\textcolor{blue}{$8.7\times10^{18}$}} & \textcolor{blue}{1.284} & -0.380 & \textcolor{black}{$1.18\times10^{13}$(A)} & \textsl{\textcolor{black}{$1.17\times10^{10}$}}\tabularnewline
xii & \textsl{\textcolor{blue}{$2.9125\times10^{20}$}} & \textcolor{black}{$3.0\times10^{20}$} & \textsl{\textcolor{blue}{$8.8\times10^{18}$}} & \textcolor{blue}{1.274} & -0.380 & \textcolor{black}{$1.18\times10^{13}$(A)} & \textsl{\textcolor{black}{$1.17\times10^{10}$}}\tabularnewline
\hline 
\end{tabular}\centering\label{SP-deep}
\end{table*}
\begin{table*}
\caption{Schrödinger\textendash Poisson calculations for various scenarios
assuming \uline{shallow} acceptor levels lying close to the valence
band edge. Values in blue have been calculated, while the rest are
fixed or predefined for each case.}
\begin{tabular}{cccccccc}
\hline 
Case & $N_{\mathrm{D}}$ & $N_{\mathrm{A}}$ & $N_{\mathrm{A}}-N_{\mathrm{D}}$ & $E_{\mathrm{C_{BULK}}}-E_{\mathrm{F}}$ & $E_{\mathrm{C_{SURF}}}-E_{\mathrm{F}}$ & $N_{\mathrm{ss}}$ & $n_{\mathrm{S}}$\tabularnewline
 & (cm\textsuperscript{-3}) & (cm\textsuperscript{-3}) & (cm\textsuperscript{-3}) & (eV) & (eV) & (cm\textsuperscript{-2}) & (cm\textsuperscript{-2})\tabularnewline
\hline 
i & \textcolor{black}{$3\times10^{17}$} & \textcolor{black}{\textemdash{}} &  & \textcolor{blue}{0.047} & -0.600 & \textsl{\textcolor{blue}{$1.18\times10^{13}$}}\textcolor{blue}{(A)} & \textsl{\textcolor{blue}{$1.33\times10^{13}$}}\textcolor{blue}{(B)}\tabularnewline
ii & \textcolor{black}{$3\times10^{17}$} & \textsl{\textcolor{blue}{$5.0\times10^{17}$}} & \textsl{\textcolor{blue}{$2.0\times10^{17}$}} & \textcolor{blue}{2.679} & -0.470 & \textsl{\textcolor{blue}{$8.73\times10^{12}$}} & \textsl{\textcolor{black}{$6.38\times10^{12}$}}\textcolor{black}{(0.48B)}\tabularnewline
iii & \textcolor{black}{$3\times10^{17}$} & \textsl{\textcolor{blue}{$6.9\times10^{17}$}} & \textsl{\textcolor{blue}{$3.9\times10^{17}$}} & \textcolor{blue}{2.697} & -0.380 & \textsl{\textcolor{blue}{$6.80\times10^{12}$}} & \textsl{\textcolor{black}{$3.46\times10^{12}$}}\textcolor{black}{(0.26B)}\tabularnewline
iv & \textsl{\textcolor{blue}{$1.61\times10^{18}$}} & \textcolor{black}{$2.0\times10^{18}$} & \textsl{\textcolor{blue}{$3.9\times10^{17}$}} & \textcolor{blue}{2.697} & -0.380 & \textsl{\textcolor{blue}{$6.80\times10^{12}$}} & \textsl{\textcolor{black}{$3.46\times10^{12}$}}\textcolor{black}{(0.26B)}\tabularnewline
v & \textsl{\textcolor{blue}{$1.961\times10^{19}$}} & \textcolor{black}{$2.0\times10^{19}$} & \textsl{\textcolor{blue}{$3.9\times10^{17}$}} & \textcolor{blue}{2.697} & -0.380 & \textsl{\textcolor{blue}{$6.80\times10^{12}$}} & \textsl{\textcolor{black}{$3.46\times10^{12}$}}\textcolor{black}{(0.26B)}\tabularnewline
vi & \textsl{\textcolor{blue}{$1.9961\times10^{20}$}} & \textcolor{black}{$2.0\times10^{20}$} & \textsl{\textcolor{blue}{$3.9\times10^{17}$}} & \textcolor{blue}{2.697} & -0.380 & \textsl{\textcolor{blue}{$6.80\times10^{12}$}} & \textsl{\textcolor{black}{$3.46\times10^{12}$}}\textcolor{black}{(0.26B)}\tabularnewline
vii & \textsl{\textcolor{blue}{$2.9961\times10^{20}$}} & \textcolor{black}{$3.0\times10^{20}$} & \textsl{\textcolor{blue}{$3.9\times10^{17}$}} & \textcolor{blue}{2.697} & -0.380 & \textsl{\textcolor{blue}{$6.80\times10^{12}$}} & \textsl{\textcolor{black}{$3.46\times10^{12}$}}\textcolor{black}{(0.26B)}\tabularnewline
viii & \textcolor{black}{$3\times10^{17}$} & \textsl{\textcolor{blue}{$2.6\times10^{18}$}} & \textsl{\textcolor{blue}{$2.3\times10^{18}$}} & \textcolor{blue}{2.743} & \textcolor{blue}{-0.540} & \textcolor{black}{$1.18\times10^{13}$(A)} & \textsl{\textcolor{black}{$3.46\times10^{12}$}}\textcolor{black}{(0.26B)}\tabularnewline
ix & \textcolor{black}{$3\times10^{17}$} & \textsl{\textcolor{blue}{$4.8\times10^{18}$}} & \textsl{\textcolor{blue}{$4.5\times10^{18}$}} & \textcolor{blue}{2.762} & -0.380 & \textcolor{black}{$1.18\times10^{13}$(A)} & \textsl{\textcolor{blue}{$7.70\times10^{9}$}}\tabularnewline
x & \textsl{\textcolor{blue}{$1.546\times10^{19}$}} & \textcolor{black}{$2.0\times10^{19}$} & \textsl{\textcolor{blue}{$4.5\times10^{18}$}} & \textcolor{blue}{2.762} & -0.380 & \textcolor{black}{$1.18\times10^{13}$(A)} & \textsl{\textcolor{black}{$7.70\times10^{9}$}}\tabularnewline
xi & \textsl{\textcolor{blue}{$1.9546\times10^{20}$}} & \textcolor{black}{$2.0\times10^{20}$} & \textsl{\textcolor{blue}{$4.5\times10^{18}$}} & \textcolor{blue}{2.762} & -0.380 & \textcolor{black}{$1.18\times10^{13}$(A)} & \textsl{\textcolor{black}{$7.70\times10^{9}$}}\tabularnewline
xii & \textsl{\textcolor{blue}{$2.9546\times10^{20}$}} & \textcolor{black}{$3.0\times10^{20}$} & \textsl{\textcolor{blue}{$4.5\times10^{18}$}} & \textcolor{blue}{2.762} & -0.380 & \textcolor{black}{$1.18\times10^{13}$(A)} & \textsl{\textcolor{black}{$7.70\times10^{9}$}}\tabularnewline
\hline 
\end{tabular}\centering\label{SP-shallow}
\end{table*}

\pagebreak\bibliographystyle{apsrev4-1}
\bibliography{SEAL_In2O3_literature}